\documentclass[letterpaper,twocolumn,10pt]{article}
\usepackage{main}

\usepackage{tikz}
\usepackage{amsmath}
\usepackage[numbers]{natbib}
\usepackage{url}
\usepackage{xcolor}
\usepackage{subfigure}
\usepackage[skip=6pt]{caption}
\usepackage{multicol}
\usepackage{tabularx}
\usepackage{booktabs}
\usepackage{graphicx}
\usepackage{pbox}
\usepackage{amsmath}
\usepackage{arydshln}
\usepackage[noend,linesnumbered,ruled]{algorithm2e}

\usepackage{breakurl}
\usepackage{float}

\usepackage[compact]{titlesec}
\titlespacing{\paragraph}{0pt}{*1}{*1}       
\titleformat{name=\section}
  {\bf\Large}{\thesection.\quad}{0pt}{}
\titleformat{name=\section,numberless}
  {\bf\Large}{}{0pt}{}[\vspace*{-2pt}]
\titleformat{\subsection}
  {\bf\large}{\thesubsection\quad}{0pt}{}
\titleformat{\subsubsection}
  {\it\large}{\thesubsubsection\quad}{0pt}{}

\usepackage{footmisc}

\def\eg{e.g.,\xspace}

\def\ie{i.e.,\xspace}

\newcommand{\name}{VirtualFlow\xspace}
\begin{document}

\date{}

\title{\Large \bf \name: Decoupling Deep Learning Models \\ [1mm] from the Underlying Hardware}

\author{
{\rm Andrew Or}\\
Princeton University
\and
{\rm Haoyu Zhang}\\
Google AI
\and
{\rm Michael J. Freedman}\\
Princeton University
} 

\maketitle

\widowpenalty10000
\clubpenalty10000

\begin{abstract}
State-of-the-art deep learning systems such as TensorFlow and PyTorch
tightly couple the model with the underlying hardware. This coupling
requires the user to modify application logic in order to run the same
job across a different set of resources, thereby limiting the choice
of hardware for a given workload and potentially forcing the user to
forgo more efficient hardware configurations.

We propose \name, a system leveraging a novel abstraction called
\textit{virtual node processing} to decouple the model from the hardware.
In each step of training or inference, the batch of input data is split across
virtual nodes instead of hardware accelerators (\eg GPUs and
TPUs). Mapping multiple virtual nodes to each accelerator and processing
them sequentially effectively time slices the batch, thereby allowing
users to reduce the memory requirement of their workloads and mimic
large batch sizes on small clusters.

Using this technique, \name enables many new use cases, such as reproducing
training results across different hardware, resource elasticity, and heterogeneous
training. In our evaluation, our implementation of \name for TensorFlow achieved
strong convergence guarantees across different hardware with out-of-the-box
hyperparameters, up to 48\% lower job completion times with resource elasticity,
and up to 42\% higher throughput with heterogeneous training.

\end{abstract}

\section{Introduction}

Modern deep learning frameworks, such as TensorFlow~\cite{tensorflow} and
PyTorch~\cite{pytorch}, make a number of simplifying assumptions about the
environment in which deep learning jobs are run today. First, a model's
convergence behavior need not be preserved across different hardware
configurations. Instead, the burden falls on the user to retune the
hyperparameters and apply custom optimization techniques in order to achieve
the same training results~\cite{facebook-imagenet, tencent-imagenet, bert-76-minutes}.
Second, resource allocations are tied to the lifetime of a job; any
adjustment to a job's allocation requires interrupting the job and
restarting it from checkpoints. Third, the set of resources allocated
to a job must be homogeneous.

\begin{figure}[t]
  \centering
  \includegraphics[width=\linewidth,trim={3cm 0.65cm 3cm 0},clip=true]{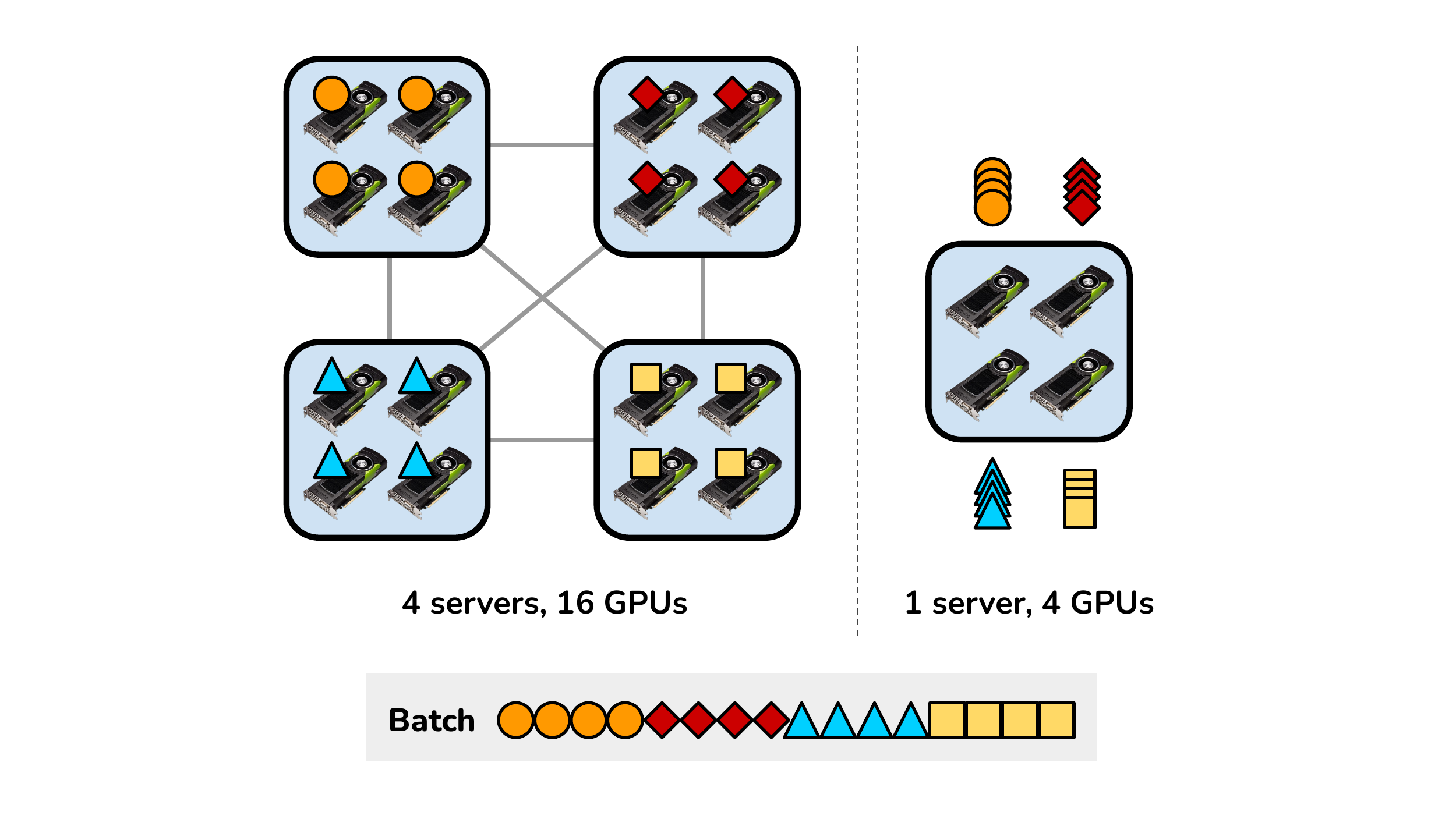}
  \caption{\textbf{Virtual node processing.} A batch is split into
    $16$~equally sized virtual nodes (colored shapes), which are distributed
    among the GPUs. Virtual nodes assigned to the same GPUs are executed
    sequentially, allowing 4 GPUs to train the same model as 16 GPUs using
    the same set of hyperparameters, including the batch size.
  }
  \vspace{-4mm}
  \label{fig:virtual-overview}
\end{figure}

In light of recent trends, however, the above assumptions fall short in
today's practical circumstances. First, the scale of deep learning workloads
continues to rise dramatically: model sizes have grown to billions of
parameters~\cite{megatron-lm, turing-nlg, gpt3}, dataset sizes to hundreds
of GBs~\cite{imagenet, t5}, and batch sizes to $64$K and above to allow for
increased parallelism~\cite{tencent-imagenet, sensetime-imagenet}.
Hardware advances have been slow to catch up, however, leading to high
computational requirements for these larger workloads. For instance,
BERT has been pre-trained on $16$~TPUs~\cite{bert} and up to $1024$~TPUs~\cite{bert-76-minutes},
and Megatron-LM, with $8.3$~billion parameters,
has been trained on $512$ V100~($32$GB) GPUs~\cite{megatron-lm}.
But given the inability to translate convergence across hardware, these results
are often unrepeatable for most researchers.

Second, shared clusters with a heterogeneous mix of GPUs are increasingly
common. For instance, Microsoft reports using a large, multi-tenant cluster
consisting of thousands of GPUs of various types shared among hundreds of
users~\cite{deep-learning-cluster-analysis}. Resource heterogeneity is also
common in small research lab settings, which often accumulate multiple
generations of GPUs over the years~\cite{gavel}.  Yet systems today are unable
to leverage this heterogeneity for individual training jobs.



\subsection{New Challenges}
\label{subsec:challenges}

These recent trends raise new important challenges or exacerbate existing ones
for today's deep learning workloads:


\textbf{High resource requirement.} Many new workloads require large clusters
of expensive hardware accelerators that are inaccessible to most users.

\textbf{Lack of experimentation.} With the increase in scale, users may wish
to experiment on a small testbed before deploying the model on a large cluster.
However, this is not possible today: the original batch size will not
fit in the memory limits of the testbed, and changing the batch size may
compromise the convergence of the model.

\textbf{Lack of reproducibility.} More generally, training results cannot be
easily reproduced across different hardware. This is because training the same
model on a different set of resources typically requires adjusting the batch
size and retuning dependent hyperparameters, such as the learning rate, in order
to achieve the same training results~\cite{facebook-imagenet, sensetime-imagenet,
tencent-imagenet, bert-76-minutes}. This is cumbersome in practice, and techniques
proposed for one workload often do not work for another~\cite{dp-survey}.

\textbf{Adapting to dynamic resource availability.} Existing attempts to
dynamically adjust a job's resource allocation must interrupt and restart the
job~\cite{gandiva, optimus, gavel}, since resource allocations are static in
today's deep learning systems. However, this is inefficient, because each
adjustment can take minutes~\cite{andrew-resource-elasticity}. Further, the
batch size may change across restarts, potentially affecting the convergence
of the model.

\textbf{Adapting to heterogeneous environments.} Today, deep learning jobs
are restricted to single types of accelerators. Being able to additionally
utilize leftover accelerators of different types can lead to faster jobs and
higher cluster utilization.

\subsection{Decoupling Model from Hardware}

All of the above challenges largely stem from a central drawback in today's
deep learning systems: \textit{a tight coupling between the model and
the underlying hardware}. This coupling comes from two main sources:

\textbf{Model graph.} Today's frameworks embed hardware configuration
information into the model graph itself. Tensor operations are explicitly
placed on specific accelerators, and communication operations involve
a fixed set of accelerators.

\textbf{Batch size.} The batch size, an important hyperparameter that has
a large effect on the convergence trajectory of a model~\cite{large-batch},
is often tied to the memory capacity of individual hardware accelerators and
the number of such accelerators in the cluster~\cite{facebook-imagenet,
tencent-imagenet, sensetime-imagenet}. If the global batch size exceeds
the cluster-wide memory limit, the workload will simply fail.

In this paper, we argue that systems-level constraints should be decoupled
from application-level semantics. A model should converge to the same accuracy
regardless of the set of resources it is trained on. Performance should degrade
gracefully with the amount of resources assigned to a job. The user should be
able to tune the model's hyperparameters once and train the model everywhere,
and the result should be the same across different hardware configurations.

The same philosophy can be observed in many big-data analytics systems.
In MapReduce-style batch processing~\cite{mapreduce, spark} and stream
processing workloads~\cite{storm, flink, spark-streaming}, the system always
computes the same answers regardless of the level of parallelism and the
amount of resources assigned to the job. The input data is sliced into many
small partitions to be processed in multiple sequential waves of tasks, and
the job would not fail if the amount of data processed in a single wave did
not fit in the aggregate memory of the system.

\subsection{Virtual Node Processing}

Towards this goal of separating the model from the hardware, this paper
introduces \textit{virtual nodes} as a substrate for distributing computation
across hardware accelerators (Figure~\ref{fig:virtual-overview}). In this paradigm,
each batch of the input data is partitioned among virtual nodes instead of hardware
accelerators. One or more virtual nodes are then mapped to each hardware accelerator
and processed sequentially on the accelerator, thus producing one or more
MapReduce-style waves of execution within each step of training or inference.

\name's approach leverages the insight that all virtual nodes share the same model
parameters. This allows the model to be cached in each accelerator's memory
at the beginning of the step and efficiently reused by all virtual nodes mapped to
that accelerator. The gradients produced by these virtual nodes are then aggregated
into a shared memory buffer on the accelerator, thus adding a small, constant overhead
independent of the number of virtual nodes used~(\S\ref{subsec:execution}).

Virtual node processing allows \name to preserve model convergence behavior across
different hardware by fixing the total number of virtual nodes, and thus the batch size
and other hyperparameters. Instead, only the mapping between virtual nodes and
hardware accelerators need to be adjusted. This enables new important use cases:

\textbf{Lower resource requirement.} Workloads that previously required large
clusters can now be packed into smaller ones by mapping many virtual nodes to
each accelerator.

\textbf{Reproducibility and experimentation} on smaller test beds is now possible,
as results obtained by other users can now be reproduced on a different set of
resources without modification of any hyperparameter or optimization strategy.

\begin{figure}
  \centering
  \setlength{\abovecaptionskip}{1mm}
  \setlength{\belowcaptionskip}{-3mm}
  \includegraphics[width=0.85\linewidth]{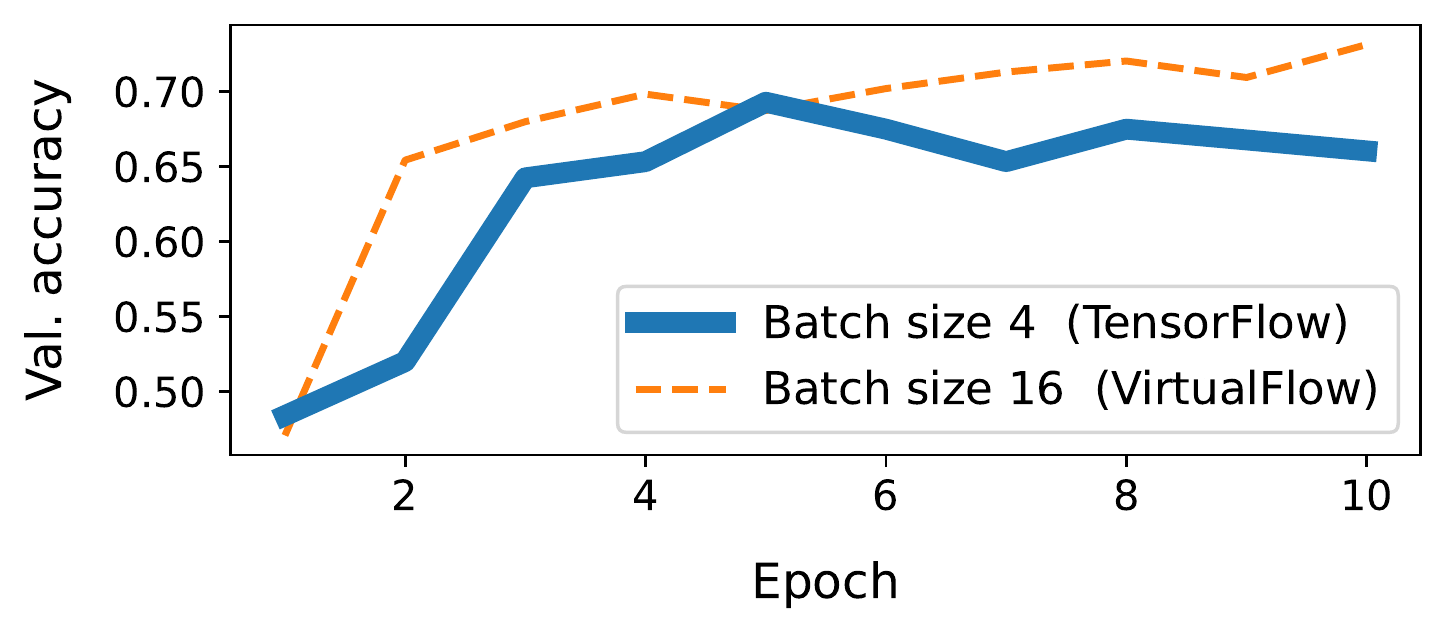}
  \caption{Training with virtual nodes (batch size 16) achieved a
    higher accuracy than was previously possible on the same set of
    resources. In this task, we fine-tune BERT-LARGE on the Recognizing
    Textual Entailment~(RTE) task on a single RTX 2080Ti GPU.}
  \label{fig:bert-large-rte-2lines}
\end{figure}

\textbf{Hyperparameter exploration.} Batch sizes previously exceeding the
aggregate limits of the underlying cluster are now accessible. In certain workloads,
being able to access these batch sizes can lead to higher model
accuracies~(Figure~\ref{fig:bert-large-rte-2lines}).

\textbf{Resource elasticity.} Dynamically resizing a job while maintaining
convergence guarantees---previously an open challenge~\cite{andrew-resource-elasticity}---is now possible.
\name achieves this by redistributing the virtual nodes among the new set of
accelerators. When scaling out, important virtual node state such as model
parameters and certain stateful kernels (\eg batch normalization variables~\cite{batch-norm})
are migrated in an all-gather operation to bootstrap the new workers~(\S\ref{subsec:redistributing}).
Unlike in state-of-the-art schedulers, the transition is seamless from the
application's perspective and the job need not be restarted.

\textbf{Heterogeneous training}---combining multiple types of accelerators in
the same job---can be expressed as distributing virtual nodes unevenly across
the accelerators, thereby assigning more data to the more powerful types.
Given a workload and a set of heterogeneous resources, \name solves for the
most efficient configuration(s) using offline profiles (\S\ref{subsec:het-assignment})
and ensures training correctness by performing weighted gradient synchronizations
(\S\ref{subsec:het-correctness}).

We implemented {\name} on top of TensorFlow and evaluated the system
on a set of representative deep learning models (ResNet~\cite{resnet},
BERT~\cite{bert}, Transformer~\cite{transformer}). To showcase the benefits
of heterogeneous training in a multi-tenant setting, we also extended Gavel~\cite{gavel}
to consider heterogeneous allocations (\S\ref{subsubsec:het-scheduler}).
In our evaluation, {\name} demonstrates strong model convergence guarantees
across different hardware, improves cluster utilization by 20\%
and reduces job completion time by 48\% with elasticity, and improves job
throughput by 42\% with heterogeneous training.

\section{Background}

In this section, we describe two important ways deep learning
workloads are tightly coupled with the underlying hardware in
state-of-the-art systems (\S\ref{subsec:hyperparameters},
\S\ref{subsec:model-graph}), then discuss the target setting
of this paper (\S\ref{subsec:setting}).

\subsection{Hyperparameters Tied to Hardware}
\label{subsec:hyperparameters}

Hyperparameters, such as the batch size, learning rate, and dropout rate,
have important effects on the convergence of a model. For this reason,
significant effort is often put into tuning these hyperparameters to
achieve desirable results.

The \textit{batch size} refers to the number of input examples, \eg images or
sentences, processed within a training or inference step. Each batch is divided
evenly among the hardware accelerators, which are assumed to be homogeneous.

Using larger batch sizes generally improves training and inference throughput. Within a single
accelerator, the \textit{local} batch size is often set to the maximum
size possible within the limits of the accelerator's memory capacity, so as to
maximize utilization of the accelerator and reduce the number of kernels
launched on it. Across multiple accelerators, the \textit{global} batch size
is simply the local batch size multiplied by the number of accelerators in
the system. Thus, the larger the global batch size, the greater the number
of accelerators that can be used to process the batch in parallel.

However, prior work has shown that large batch sizes tend to
deteriorate model convergence~\cite{large-batch}. In order to preserve
convergence behavior while scaling a workload, various efforts have
proposed to adjust hyperparameters dependent on the batch size, such
as the learning rate~\cite{facebook-imagenet}, or even to apply custom
optimization algorithms~\cite{sensetime-imagenet, tencent-imagenet, lars, bert-76-minutes}.

\textbf{Hurdles for reproducibility.}
Thus, reproducing existing results on a different set of hardware requires
significant effort and expertise. In some cases, it is even impossible. For
example, the results from training the the BERT model using a batch size of
$32,000$ examples on $1,024$~TPUs~\cite{bert-76-minutes} and
$1,472$~GPUs~\cite{bert-nvidia} cannot be reproduced on a smaller test bed of
$16$~GPUs, as the same batch size would not fit in the smaller cluster's GPU
memory. On the other hand, reducing the batch size would inevitably lead to
very different convergence trajectories that require retuning various
hyperparameters. This poses a major hurdle for experimentation as well as
scaling.

\subsection{Inflexible Model Graph}
\label{subsec:model-graph}

Another source of coupling between the model and the hardware lies in the
\textit{model graph}, which specifies the network of operations to perform
on the input data. Modern deep learning frameworks compile and optimize this
graph once at the beginning of training and reuse it for the rest of the job.

In addition to tensor operations, information regarding the underlying cluster
configuration is also embedded into the model graph. In both TensorFlow and
PyTorch, for instance, the graph is defined under a \textit{distribution
strategy} that specifies how model parameters should be synchronized in
different settings, such as single GPU, single machine multi-GPU, and
distributed multi-GPU.

\textbf{Hurdles for resource elasticity.}
Once the model graph is created under a particular distribution strategy,
subsequent training will use synchronization operations that involve a
fixed set of hardware accelerators. Adjusting a job's resource allocation
would involve rebuilding the entire graph under a new distribution strategy
and reloading previously trained model parameters from a checkpoint, an
expensive process that can take minutes~\cite{andrew-resource-elasticity}.
Further, as discussed in~\S\ref{subsec:hyperparameters}, changing the
amount of resources in the middle of a job can lead to adverse effects
on the model's convergence.

\subsection{Data Parallel, Synchronous Training}
\label{subsec:setting}

The most common form of parallelism in distributed deep learning workloads
is \textit{data parallelism}, where the model graph is replicated across
the hardware accelerators, and each accelerator processes its share of the
batch independently. This is in contrast to \textit{model parallelism}, which
partitions the model graph across the accelerators instead, and is used
primarily for extremely large models that do not fit in the memory of a
single accelerator.

In modern workloads, data parallelism is typically combined with
\textit{synchronous training}, which enforces a synchronization barrier
at the end of each step, and is shown to have better convergence properties
than \textit{asynchronous training}~\cite{revisiting-sync}. Gradients can
be synchronized using either the parameter server architecture~\cite{parameter-server}
or the all-reduce architecture~\cite{allreduce, horovod}, though the latter
is increasingly common.

This paper targets data parallel, synchronous training, though many of
the techniques proposed are also applicable to the model parallel setting.
This is explored further in~\S\ref{sec:future}.

\section{Virtual Node Processing}
\label{sec:virtual}

\begin{figure}[t]
  \centering
  \setlength{\abovecaptionskip}{0mm}
  \setlength{\belowcaptionskip}{-1mm}
  \vspace{-3mm}
  \includegraphics[width=0.9\linewidth, trim={-0.75cm 10cm 13.5cm 0}]{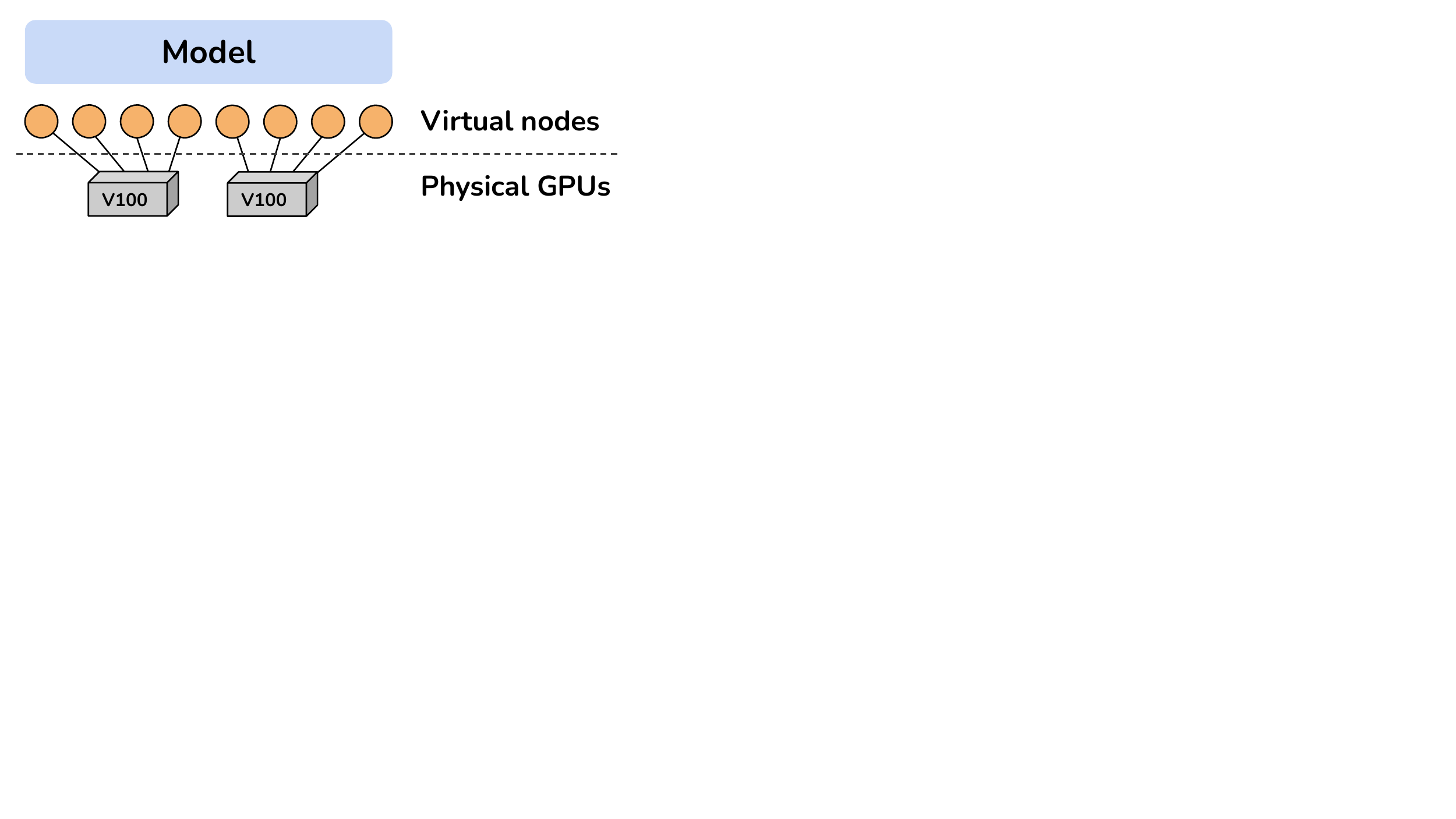}
  \caption{Mapping between virtual nodes and hardware accelerators
    is flexible, but only virtual nodes affect model convergence. Thus,
    changes to resource allocations are hidden from the application.}
  \vspace{-3mm}
  \label{fig:mapping}
\end{figure}

The core concept in \name is virtual node processing, a layer of indirection
between the model and the hardware. From the model's perspective, \textit{virtual
nodes}, rather than hardware accelerators, perform the computation.
As long as the total number of virtual nodes is unchanged, the batch size and 
thus the convergence properties of the model also remain the same.
Then, flexibility in resource allocation comes from adjusting the mapping
between the virtual nodes and the underlying hardware accelerators, which
have no effect on model training semantics (Figure~\ref{fig:mapping}).

\subsection{Time and Resource Trade-off}
\label{subsec:tradeoff}

\begin{figure}[t]
  \centering
  \setlength{\belowcaptionskip}{0mm}
  \vspace{-3mm}
  \includegraphics[width=0.85\linewidth,trim={0cm 6cm 11cm 0.25cm}]{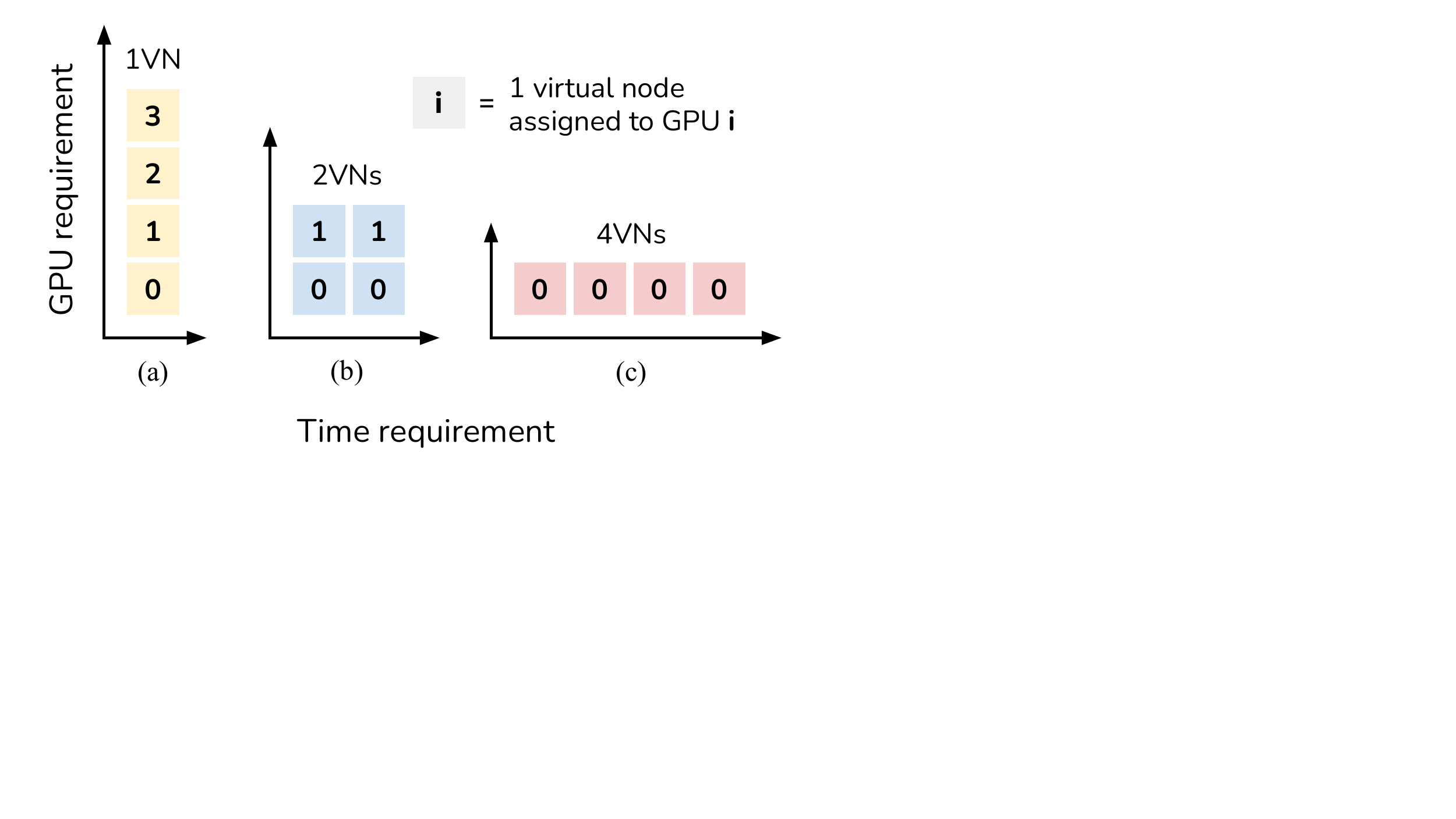}
  \vspace{-2mm}
  \caption{Virtual node trade-off between resource requirement
    and time requirement. VN in this figure refers to number of
    virtual nodes assigned to each hardware accelerator. The design
    space for today's deep learning workloads is limited to only (a).}
  \vspace{-3mm}
  \label{fig:virtual-tradeoff}
\end{figure}

Today's deep learning systems are a special case of virtual node processing
that uses one virtual node per hardware accelerator (Figure~\ref{fig:virtual-tradeoff}a).
However, this is only one possible configuration in the trade-off space
between time and resource requirements. \name divides the compution in each
batch in the time dimension as well as the spatial dimension
(Figure~\ref{fig:virtual-tradeoff}b and c), processing the virtual nodes
assigned to the same accelerators sequentially. This provides users with
the freedom to gracefully fall back to running on fewer accelerators with
longer training times.

This flexibility is crucial to model reproducibility, experimentation,
and hyperparameter exploration. By preserving convergence behavior across
different hardware configurations (\eg Figure~\ref{fig:virtual-tradeoff}a, b, and c),
\name allows users to replicate training results produced by others
regardless of the resources used. Experimentation on smaller testbeds
is now possible by using many virtual nodes on each hardware accelerator
to mimic the larger deployment. On the other hand, users can explore the
effects of using previously inaccessible batch sizes on the same set of
resources by increasing the number of virtual nodes used on each accelerator.

\begin{figure*}[t]
  \centering
  \setlength{\belowcaptionskip}{-1mm}
  \includegraphics[width=0.95\linewidth,trim={0 4cm 0 3cm}]{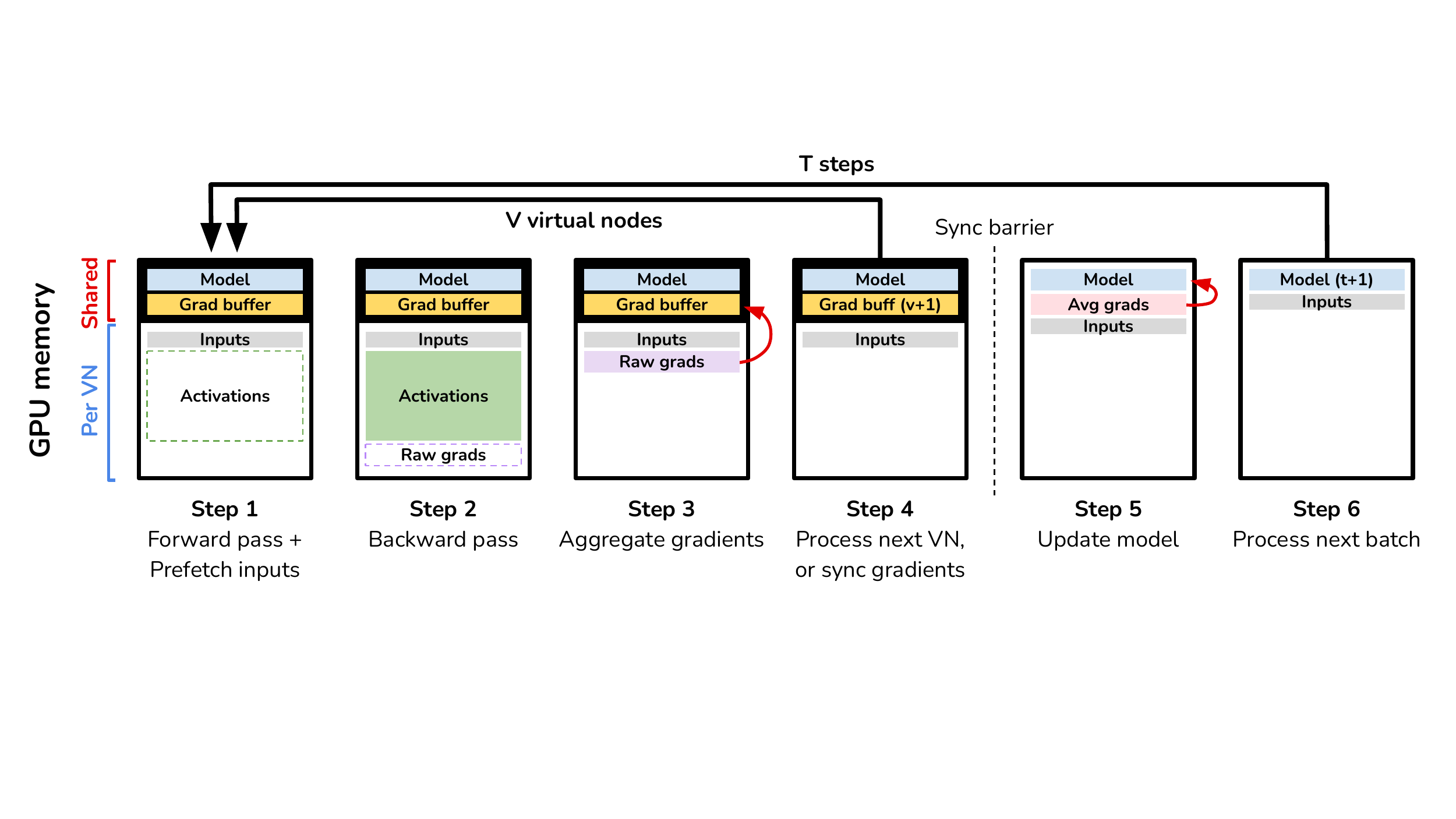}
  \vspace{0mm}
  \captionsetup{width=0.95\textwidth}
  \caption{
    Memory footprint of virtual node processing in a single training step.
    The model parameters and the gradient buffer are shared across all virtual nodes
    on each accelerator, while the memory used in each forward and backward pass is
    specific to individual virtual nodes. Memory overhead from the gradient buffer
    is a small constant independent of the number of virtual nodes $V$ per accelerator.
  }
  \label{fig:memory-cycle}
\end{figure*}

\subsection{Virtual Node Execution}
\label{subsec:execution}

Each batch of the input data is split among the virtual nodes in a manner
analogous to how a job in MapReduce is divided into tasks. Virtual nodes
assigned to the same hardware accelerator are processed sequentially, while
virtual nodes assigned to different accelerators are still processed in
parallel. This produces one or more \textit{waves} of execution, similar to
MapReduce workloads where the number of tasks is often a small multiple of
the number of CPU slots in the system.

Figure~\ref{fig:memory-cycle} traces the steps involved in processing a single
batch of data with virtual node processing. In each training step, \name computes
$V$ forward and backward passes, where $V$ is the number of virtual nodes on each
accelerator. During the forward pass, \name computes the activations while
prefetching inputs for the next virtual node in the background~(Step~1).
At the end of the backward pass~(Step~2), local gradients are aggregated into a
gradient buffer shared across all virtual nodes on the accelerator~(Step~3).
When there are no more virtual nodes to process, the locally aggregated gradients
are synchronized across the cluster~(Step~4) and each accelerator independently
applies the averaged gradients to its copy of the model as before~(Step~5).

An important insight of \name's approach is that the model can be cached in
memory to be efficiently reused in all $V$ forward passes. If there is only a
single virtual node per accelerator ($V = 1$), the system falls back to prior
behavior.

\subsection{Memory Overhead}
\label{subsec:memory-overhead}

\begin{figure}
  \centering
  \setlength{\belowcaptionskip}{-4.5mm}
  \includegraphics[width=0.85\linewidth, trim={0 0 0 0.75cm}]{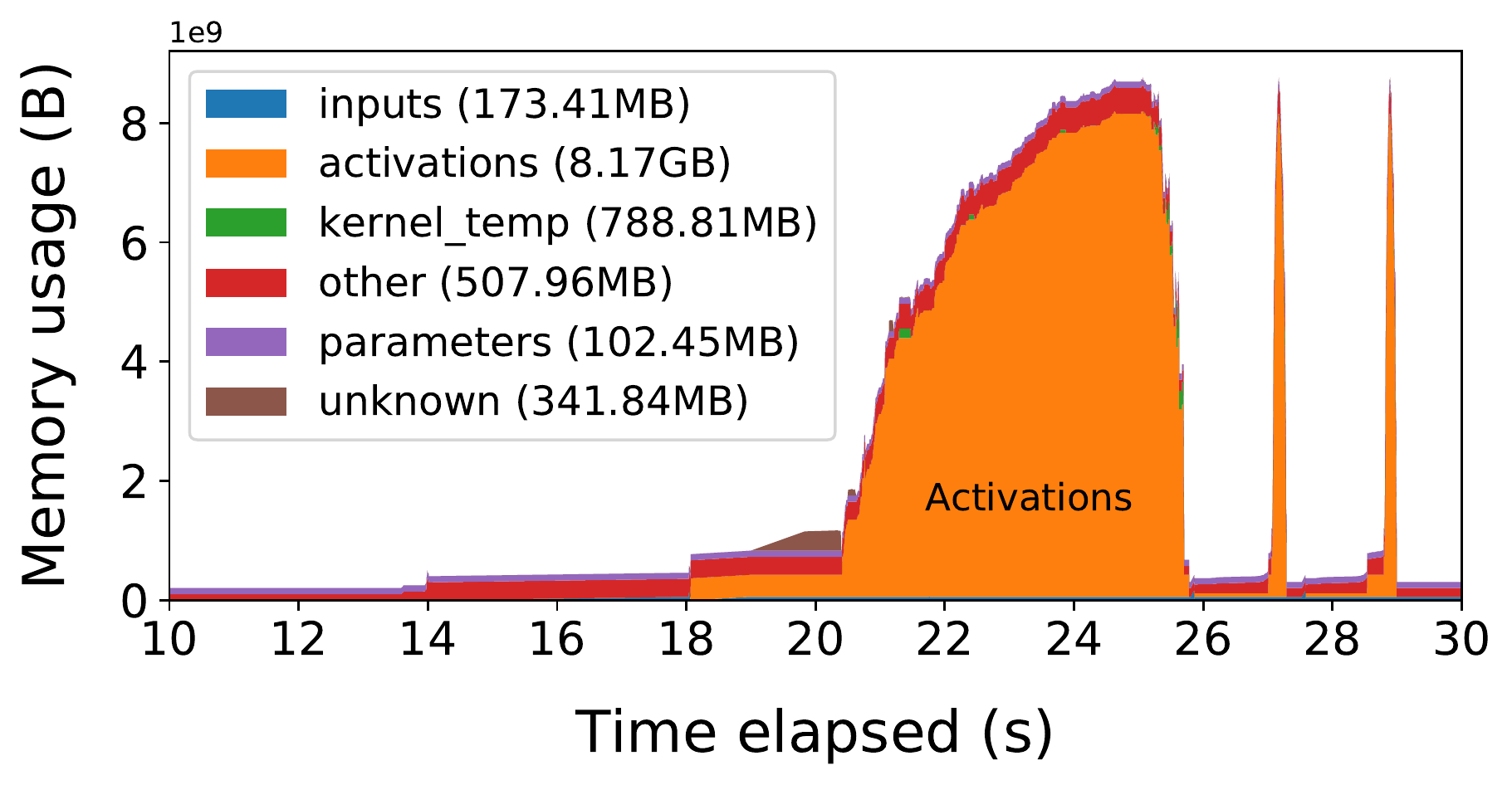}
  \caption{Memory usage in the first 3 steps of training ResNet-50 on ImageNet~\cite{imagenet}
    on a single 2080 Ti GPU, broken down by category. Activations constitute the
    vast majority of memory usage at the peak. The first step is slower due to
    initial graph optimizations.}
  \label{fig:resnet-memory-breakdown}
\end{figure}

The gradient buffer in \name is a source of memory overhead (Figure~\ref{fig:memory-cycle}).
However, because this buffer is shared among all virtual nodes assigned to the
same accelerator, the memory overhead is a \emph{constant} independent of the
number of virtual nodes on the accelerator.

The size of this buffer is the same as the model, which is a small fraction of
the peak memory usage for most workloads. Instead, memory usage is typically
dominated by activations computed during the forward pass, which scale with the
batch size while the model does not. For example, when training ResNet-50 on
ImageNet, the activations typically require over $8$GB, while the model is only
around $104$MB (see Fig.~\ref{fig:resnet-memory-breakdown}).

\section{Resource Elasticity}
\label{sec:elasticity}

Jobs that run on shared clusters can benefit significantly from resource
elasticity. This enables jobs to adapt their resource usage to changing
allocations from the cluster scheduler, which may perform such
adjustments to enforce fairness~\cite{locality2, zaharia2010delay},
preemption~\cite{pdq}, and utility-based~\cite{slaq} scheduling policies.

Elasticity has been a desireable feature in many other distributed systems,
including ones from batch processing~\cite{spark-autoscaling}, stream
processing~\cite{stream-autoscaling}, cluster
management~\cite{kubernetes-autoscaling}, and cloud
computing~\cite{aws-autoscaling, azure-autoscaling, gce-autoscaling},
with important benefits such as higher cluster utilization
and lower job completion time. In this section, we describe how \name
can bring the same benefits to distributed deep learning workloads by
expressing elasticity in terms of redistributing virtual nodes across
accelerators.

\subsection{Redistributing Virtual Nodes}
\label{subsec:redistributing}

\name maintains a mapping between virtual nodes and hardware accelerators,
but this mapping need not be fixed over time. To enable resource elasticity,
virtual nodes can be redistributed dynamically across the accelerators
assigned to a job in response to cluster demand.

More specifically, downsizing a job can be expressed in terms of moving
virtual nodes from released hardware accelerators to the remaining ones
that are still allocated to the job. Similarly, upsizing a job can be expressed
in terms of migrating a subset of the virtual nodes assigned to existing
accelerators to the new accelerators that were added. In both cases, the
total number of virtual nodes remains the same, and so adjustments to
a job's resource allocation are seamless from the perspective of the
application.

When scaling out, certain virtual node state must be migrated to the new
accelerators, including the model parameters and certain \textit{stateful kernels}.
One example of the latter is the batch normalization moving mean and
variance~\cite{batch-norm}, which are computed independently on each
accelerator and never synchronized. Bootstrapping new workers without
also migrating these stateful kernels would effectively reset their
internal state, potentially hurting convergence. \name migrates these
stateful kernels as well as the model parameters through an all-gather
operation performed on the new workers. This process typically takes
less than a second (similar to all-reduce) and only takes place once per
resource adjustment.

Figure~\ref{fig:virtual-overview} illustrates an example of resizing
a job from 16 GPUs to 4 GPUs. Each batch is split into 16 equally sized
virtual nodes, which were all assigned to different GPUs initially.
The virtual nodes are then redistributed among the 4 remaining GPUs,
such that each GPU is assigned 4 virtual nodes (instead of 1) in the
new configuration.

\begin{figure*}[t]
  \centering
  \setlength{\abovecaptionskip}{0mm}
  \setlength{\belowcaptionskip}{-3mm}
  \subfigure{\includegraphics[width=0.77\textwidth,trim={0 8.5cm 0 0}]{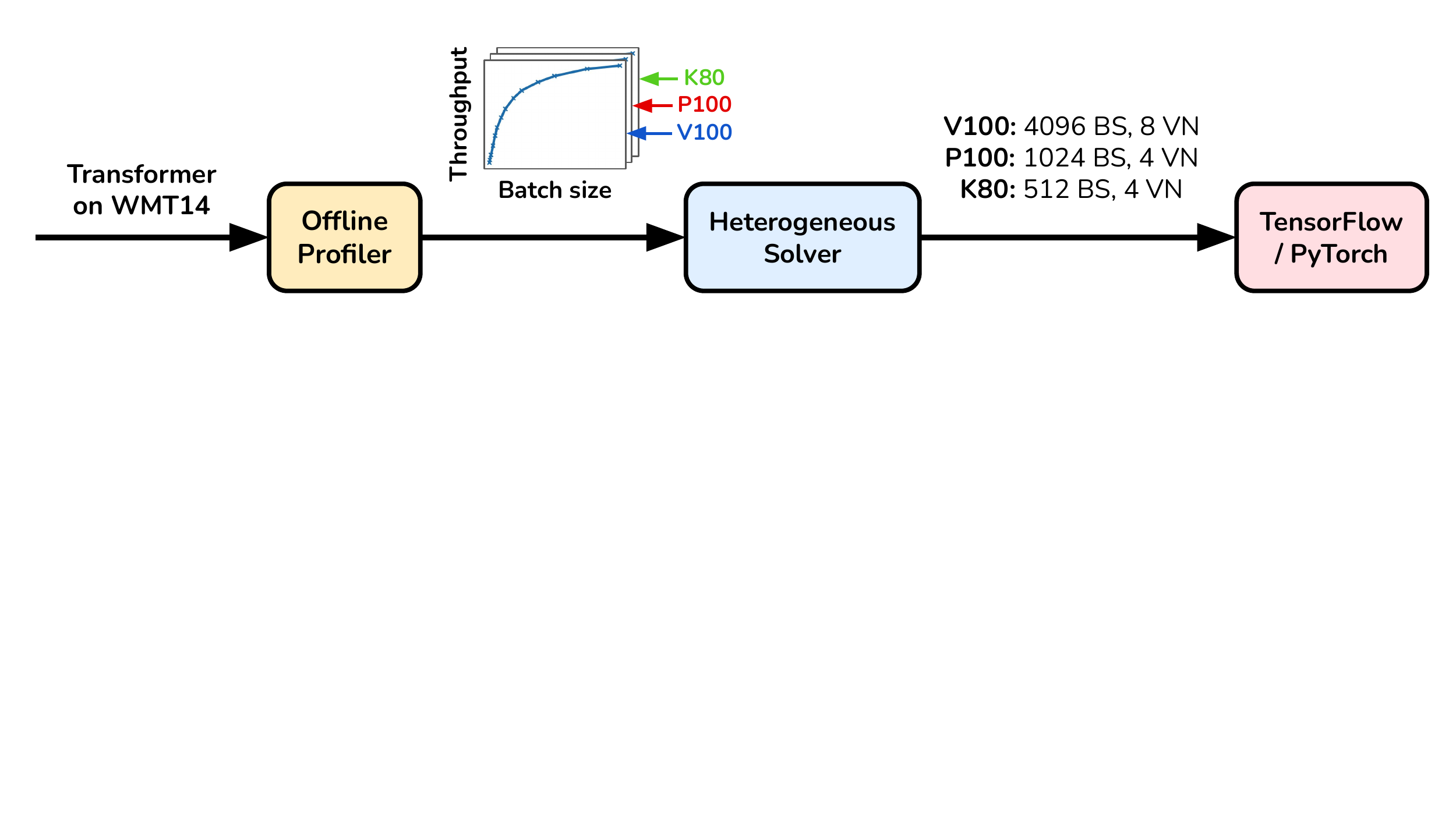}}
  \hspace{3mm}
  \subfigure{\includegraphics[width=0.2\textwidth]{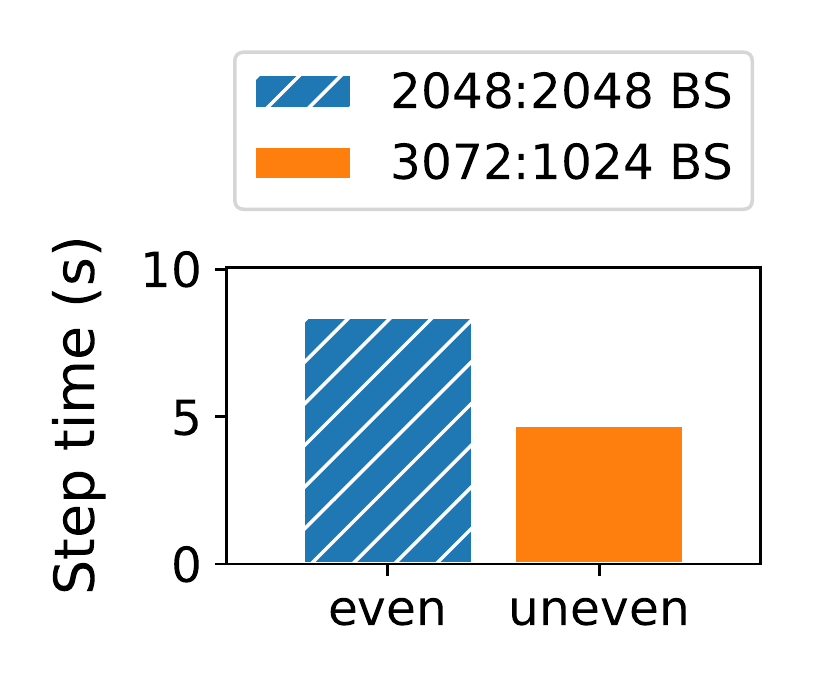}}
  \captionsetup{width=0.95\textwidth}
  \caption{(Left) \name heterogeneous training overview. (Right) Splitting
    a batch evenly across a set of uneven resources is inefficient. In this
    workload, we wish to train ResNet-50 on ImageNet on 2 V100 GPUs and 2
    P100 GPUs with a global batch size of 8192. "3072:1024 BS" means 3072
    examples are assigned to each V100 GPU, while 1024 are assigned to each
    P100 GPU per batch. In this example, the heterogeneous solver will output
    the more efficient uneven configuration.}
  \label{fig:het-overview}
\end{figure*}

\subsection{Elastic Weighted Fair Sharing (WFS)}
\label{subsec:wfs-scheduler}

To showcase the benefits of expressing elasticity in terms of
virtual nodes, we built a simple event-driven cluster scheduler
that dynamically resizes deep learning jobs based on their
relative weighted fair shares (WFS)~\cite{wfq}.
These fair shares are computed based on the priority of the jobs,
which can be set to arbitrary attributes of the job to express a
variety of scheduling objectives, such as Shortest Job First (SJF)
and Shortest Remaining Time First (SRTF). The main scheduling
logic is summarized in Algorithm~\ref{alg:wfs-scheduler}.

\begin{algorithm}[t]
  \small
  \DontPrintSemicolon
  \SetKw{Break}{break}
  \underline{function schedule} $(running\_jobs, job\_queue)$:\;
  $new\_allocations$ = expand current allocations\;
  \While{$job\_queue$ \normalfont{not empty}}{
    $fair\_allocations$ = compute fair shares(\
      $running\_jobs$, $job\_queue.peek()$)\;
    \eIf{\normalfont{no higher priority job allocations are affected}}{
      $new\_allocations$ = $fair\_allocations$\;
      $running\_jobs \hspace{1mm} \texttt{+=} \hspace{1mm} job\_queue.dequeue()$\;
    }{
      \Break\;
    }
  }
  resize jobs($new\_allocations$)\;
  \caption{Elastic WFS Scheduler}
  \label{alg:wfs-scheduler}
\end{algorithm}
\setlength{\textfloatsep}{8pt}

This scheduler has two important distinctions from existing GPU
cluster schedulers~\cite{gandiva, optimus, gavel}.
First, resource adjustments need not interrupt the jobs and restart
them from checkpoints. Second, while existing schedulers dynamically
adjust cluster-wide resource allocations, they are unable to resize
individual jobs without potentially hurting model convergence. As an
example, suppose job A demands 2 GPUs and job B demands 8 GPUs, and
there are no other jobs in the queue. If there are 8 GPUs in total,
jobs A and B will never be able to run at the same time, and 6 GPUs
will have to go idle for the entire duration of job A. Regardless of
the scheduling policy, being able to dynamically adjust the resource
requirement of each job will generate many new scheduling opportunities,
potentially leading to higher cluster utilization~(\S\ref{subsec:elasticity}).

\section{Heterogeneous Training}
\label{sec:het}

An important assumption made by state-of-the-art frameworks is resource
allocations must be \textit{homogeneous}. However,
this idealistic assumption is not well-suited for today's clusters with
multiple generations of hardware accelerators, which are increasingly
common~\cite{gavel, deep-learning-cluster-analysis}. Similarly, in a
research lab setting, users may only have access to a scarce, assorted
collection of various accelerator types. However, today's deep learning
jobs are limited to individual accelerator types and must leave unused
accelerators of different types idle.

\name relaxes this assumption by allowing users to combine multiple
accelerator types in the same job, potentially leading to significant
improvements in job throughput and cluster utilization. There are
two main challenges involved, however. First, how to distribute virtual
nodes across heterogeneous resources efficiently~(\S\ref{subsec:het-assignment})?
Second, how to provide the same semantics as homogeneous
training~(\S\ref{subsec:het-correctness})?

\subsection{Virtual Node Assignment}
\label{subsec:het-assignment}

The key intuition is to assign more virtual nodes to the resource
types with higher compute capabilities, so as to balance the step
times across the different accelerator types. This allows users to
scale their workloads by reducing the amount of computation
required on the each set of homogeneous accelerators, thereby
improving the overall throughput.

For additional flexibility in deciding how to split the batch across
the heterogeneous set of resources, we further relax the assumption
in \name that the size of each virtual node must be the same across
all accelerators. Then, determining an efficient assignment
of virtual nodes involves two steps, as outlined in
Figure~\ref{fig:het-overview} (left). First, \name performs an offline
profile of the given workload on all target accelerator types. Then,
using these offline profiles, \name solves for a configuration that
minimizes the overall step time.

\subsubsection{Offline Profile}
\label{subsubsec:het-profile}

To generate an offline profile, \name runs the given workload on a
single hardware accelerator at a time across all batch sizes of interest
that fit in the accelerator's memory. Due to memory alignment, we only consider
batch sizes that are powers of 2 or power-of-2-like numbers (\eg 48,
192, 768), which are the mid-points between adjacent powers of 2.
The process is then repeated on all accelerator types and the result is a
set of throughput over batch size curves
(Figure~\ref{fig:het-overview}, left), one for each accelerator type.


For each batch size, we only need to run a few steps (\eg 20) to
arrive at a representative average throughput, since the performance
is typically consistent across steps. Therefore, the entire process
typically takes no longer than 10 minutes, a small fraction of the
job duration for many deep learning workloads, which can run for
many hours or even days.

\subsubsection{Heterogeneous Solver}
\label{subsubsec:het-solver}

To understand why a solver is necessary, consider the scenario in
Figure~\ref{fig:het-overview} (right), in which we are given 2 V100
GPUs and 2 P100 GPUs, all with the same memory capacity of 16GB.
A naive, even split of the batch would assign the same amount of
data and the same number of virtual nodes to each GPU, regardless of
the GPU type. However, this configuration is inefficient, because, for
this workload, V100 GPUs are 4x as fast as P100 GPUs, so the system
will be bottlenecked on the P100 GPUs, leaving the V100 GPUs idle
for a large fraction of the training time. Instead, an uneven split
that assigns more data in each batch to the V100 GPUs will result
in a much shorter (44\%) overall step time.

In order to arrive at an appropriate split across the different
accelerator types, we formulate the problem as follows. For simplicity,
we treat the resources as GPUs:

\setlength{\abovedisplayskip}{-3pt}
\setlength{\belowdisplayskip}{-3pt}
\setlength{\abovedisplayshortskip}{-3pt}
\setlength{\belowdisplayshortskip}{-3pt}
\begin{align*}
  \text{Objective} \quad & \min{\max_i{(t_i(b_i) \cdot v_i + comm)}}\\
  \text{Constraint} \quad & \sum_i{n_i \cdot b_i} = B\\
  \text{Solve for} \quad & b_i, v_i, n_i \quad \forall i
\end{align*}

\begin{align*}
  B &= \text{Global batch size}\\
  b_i &= \text{Per GPU batch size for GPU type $i$}\\
  v_i &= \text{Num virtual nodes on each GPU of type $i$}\\
  t_i(b_i) &= \text{Step time on GPU type $i$}\\
  n_i &= \text{Number of GPUs of type $i$}\\
  comm &= \text{Communication overhead}\\
\end{align*}

\noindent This objective aims to equalize the step times across all GPU types
so as to minimize the overall step time, which is bottlenecked by
the slowest GPU. We multiply the step time by the number of virtual
nodes $v_i$ to reflect the fact that virtual nodes on a given GPU
are executed sequentially. The step times for each GPU type $t_i$
are constants supplied by offline profiles computed previously.
The communication overhead $comm$ can be estimated as part of the offline
profile by taking the difference between the distributed and single node
step times, using synthetic data and a local batch size of 1 in both cases
to isolate the time spent on gradient synchronization.

The solver falls back to recommending homogeneous allocations when
there are no heterogeneous combinations that can improve the throughput
of the job. This can happen if the compute capabilities are vastly
different across the GPU types, and there are not enough of the
slower GPUs to compensate for the discrepancy in performance.

\subsection{Correctness}
\label{subsec:het-correctness}

An important goal of \name is to preserve training semantics
regardless of the underlying hardware configuration. Maintaining
this abstraction for heterogeneous training involves solving two
main challenges:

\textbf{Gradient synchronization.} Existing implementations of
gradient synchronzation first take the average of the gradients
computed on the local batch, then take the average of these local
averages across all hardware accelerators. On heterogeneous resources,
however, this method can produce incorrect gradients. For instance,
suppose we assign 6 input examples to GPU0 and 2 input examples
to GPU1 in each batch. Taking a simple average will result in:

{
  \small
  \vspace{-3mm}
  \begin{align*}
    \frac{1}{2}\left(\frac{g_1 + ... + g_6}{6}\right) + \frac{1}{2}\left(\frac{g_7 + g_8}{2}\right) = \frac{g_1 + ... + g_6 + 3(g_7 + g_8)}{12}
  \end{align*}
  \vspace{-3mm}
}

\noindent where the gradients on GPU1 are weighed
disproportionately compared to the rest. Instead, \name performs
a \textit{weighted average} during gradient synchronization:

{
  \small
  \vspace{-3mm}
  \begin{align*}
    \frac{3}{4}\left(\frac{g_1 + ... + g_6}{6}\right) + \frac{1}{4}\left(\frac{g_7 + g_8}{2}\right) = \frac{g_1 + ... + g_8}{8}
  \end{align*}
  \vspace{-3mm}
}

\noindent This ensures all gradients are considered equally
regardless of how the data is distributed across the accelerators.

\textbf{Data sharding.} Similarly, existing sharding techniques
assume the batch is split evenly across the accelerators. Naively
reusing these techniques for heterogeneous training will result
in certain input examples being observed more often than others.
\name maintains the exactly-once data semantics of homogeneous
training by sharding the dataset unevenly to match the relative
local batch sizes (\eg 4:1) across the different accelerator types.

\section{Evaluation}
\label{sec:eval}

We implemented \name with resource elasticity and heterogeneous training
support on top of TensorFlow $2.4$ in $2700+$ lines of code. For elasticity,
we used the same mechanisms as~\cite{andrew-resource-elasticity}, in which
Horovod~\cite{horovod} was used as the narrow waist communication layer
that connects a changing set of worker processes. In this section, we
evaluate \name's effectiveness in reproducing results across different
hardware (\S\ref{subsec:reproducibility}), exploring previously inaccessible
hyperparameters (\S\ref{subsec:exploration}), providing elasticity while
preserving model semantics (\S\ref{subsec:elasticity}), and enabling
heterogeneous training (\S\ref{subsec:het}).

\subsection{Experimental Setup}
\label{subsec:experimental-setup}

End-to-end reproducibility and elasticity experiments are performed
on 2 servers, each with 8 NVIDIA V100 GPUs (16GB), 64 Intel Xeon CPUs
(2.2Ghz), and 250GB of DRAM, connected over a 16 Gbps connection.
Heterogeneous training experiments additionally use 2 extra similar
servers, each with 4 NVIDIA P100 GPUs (16GB). Exploration and microbenchmark
experiments use 2 NVIDIA GeForce RTX 2080Ti GPUs on a server with
32 Intel(R) Xeon(R) E5-2620v4 CPUs (2.1GHz) and 64GB of DRAM.

\begin{table}[t]
  \small
  \centering
  \setlength\dashlinedash{0.75pt}
  \setlength\dashlinegap{2pt}
  \setlength\arrayrulewidth{0.75pt}
  \setlength{\abovecaptionskip}{4mm}
  \setlength{\belowcaptionskip}{0mm}
  \begin{tabular}{c|ccc|cc}
    \noalign{\hrule height 1.5pt}
    & \multicolumn{3}{c|}{\name} & \multicolumn{2}{c}{TF*} \\
    GPUs & BS & VN$_{GPU}$ & Acc (\%) & BS & Acc (\%)\\
    \noalign{\hrule height 1pt}
    1 & 8192 & 32 & 75.92 & 256 & 69.25\\
    2 & 8192 & 16 & 75.96 & 512 & 67.30\\
    4 & 8192 & 8 & 75.99 & 1024 & 70.68\\
    8 & 8192 & 4 & 75.83 & 2048 &73.04\\
    16 & 8192 & 2 & 75.68 & -- & --\\
    2$^{\dagger}$ & 8192 & 32 & 76.01 & -- & --\\
    \hdashline
    Target & 8192 & -- & 76.26~\cite{facebook-imagenet} & -- & --\\
    \noalign{\hrule height 1.5pt}
  \end{tabular}
    \caption{\textbf{Reproducibility:} Final top-1 validation accuracies
    for the same ResNet-50 experiment shown in Figure~\ref{fig:resnet-imagenet-V100}.
    \name preserves the target accuracy of 76\% ($\pm$ 0.5\%) regardless of
    the number of GPUs assigned, while the naive solution (TF*) diverges.
    VN$_{GPU}$ refers to number of virtual nodes per GPU, and
    $\dagger$ refers to training on RTX 2080Ti GPUs instead of on V100 GPUs.}
  \label{tab:resnet-imagenet-V100}
\end{table}

\begin{table}[t]
  \small
  \centering
  \setlength\dashlinedash{0.75pt}
  \setlength\dashlinegap{2pt}
  \setlength{\abovecaptionskip}{4mm}
  \setlength{\belowcaptionskip}{0mm}
  \begin{tabular}{cccccc}
    \noalign{\hrule height 1.5pt}
    \multicolumn{3}{c}{} & QNLI & SST-2 & CoLA \\
    GPUs & BS & VN$_{GPU}$ & Acc (\%) & Acc (\%) & Acc (\%) \\\noalign{\hrule height 1pt}
    1 & 64 & 8 & 90.86 & 92.07 & 83.01\\
    2 & 64 & 4 & 91.05 & 92.35 & 84.08\\
    4 & 64 & 2 & 90.86 & 92.20 & 83.50\\
    8 & 64 & 1 & 90.88 & 91.86 & 82.45\\
    \hdashline
    Target & 64 & -- & 90.90 & 91.97 & 82.36\\
    \noalign{\hrule height 1.5pt}
  \end{tabular}
  \caption{\textbf{Reproducibility}: Final top-1 validation accuracies
    achieved by \name for fine-tuning BERT-BASE across 3 GLUE tasks.
    \name was able to reproduce the same training results as the
    state-of-the-art on a variety of hardware configurations. Previously,
    a batch size of 64 would not fit in the memory of 1 V100 GPU.}
  \label{tab:bert-base-V100}
\end{table}

\subsection{Reproducibility}
\label{subsec:reproducibility}

We demonstrate \name can reproduce training results across different
cluster sizes for two well-known deep learning workloads: ResNet-50~\cite{resnet}
on ImageNet~\cite{imagenet} and BERT~\cite{bert} fine-tuning on GLUE~\cite{glue}.
We varied the number of GPUs from 1 to 16 (8 for BERT) while fixing
the global batch sizes, and observed almost identical convergence
trajectories across different allocations for both workloads.

\textbf{Baseline.} We compare \name with a version of vanilla TensorFlow
that does not retune hyperparameters across batch sizes. For example,
for ResNet, we do not apply the linear scaling rule~\cite{facebook-imagenet}
to tune the learning rate when attempting to simulate large workloads on
smaller sets of GPUs. This setup is motivated by the fact that these
optimization techniques are often workload-specific and difficult to
identify for arbitrary workloads~\cite{dp-survey}.



\begin{figure}[t]
  \centering
  \setlength{\belowcaptionskip}{-3mm}
  \includegraphics[width=0.8\linewidth]{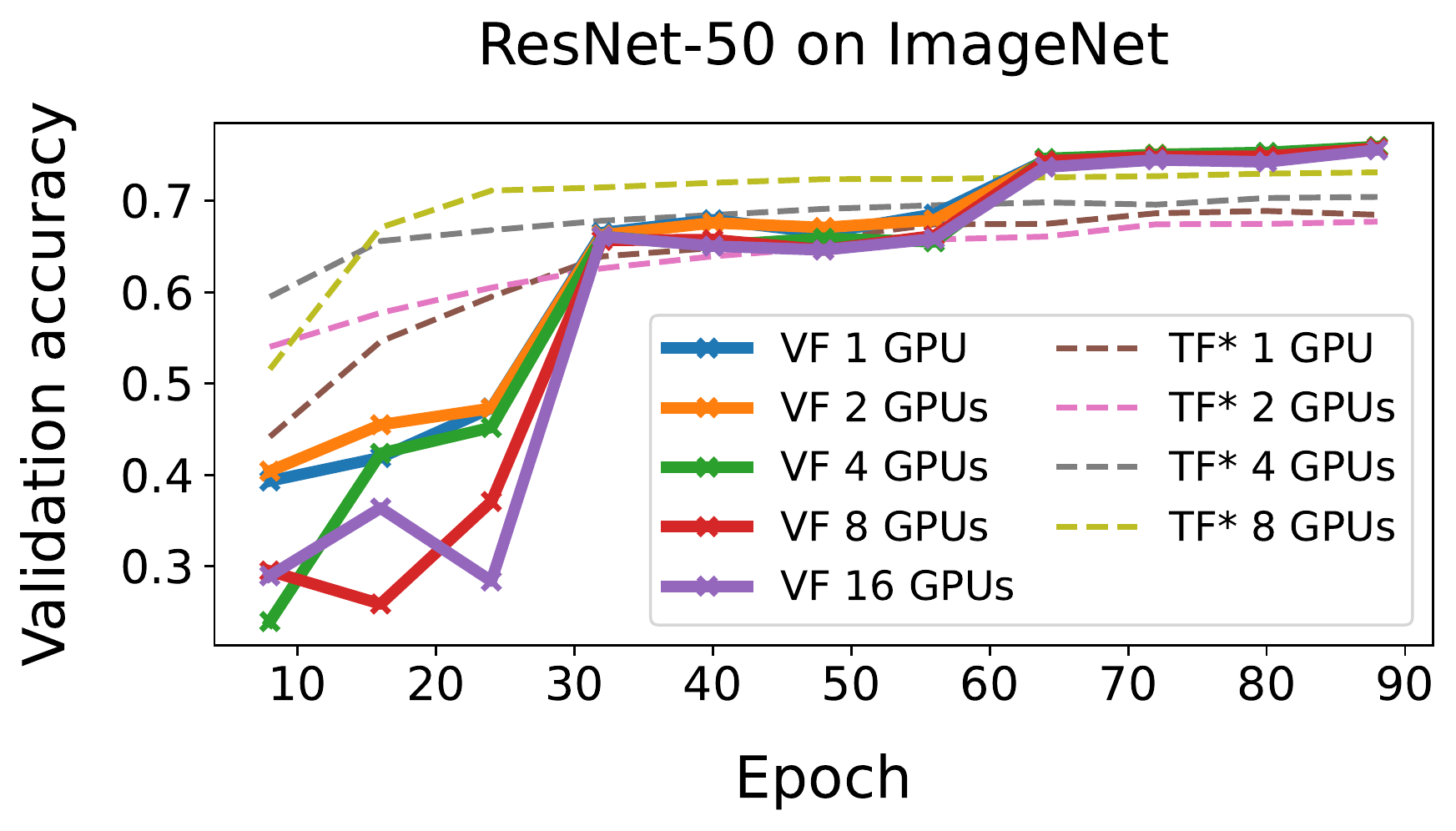}
  \caption{\textbf{Reproducibility:} \name preserves the convergence
    trajectory across different numbers of GPUs by fixing the batch
    size at 8192. Naively attempting to reproduce the same workload
    on fewer GPUs without retuning the hyperparameters (TF*) yields
    lower accuracies and different convergence behavior.}
  \vspace{3mm}
  \label{fig:resnet-imagenet-V100}
\end{figure}

\begin{figure*}[t]
  \centering
  \subfigure{\includegraphics[width=0.26\textwidth]{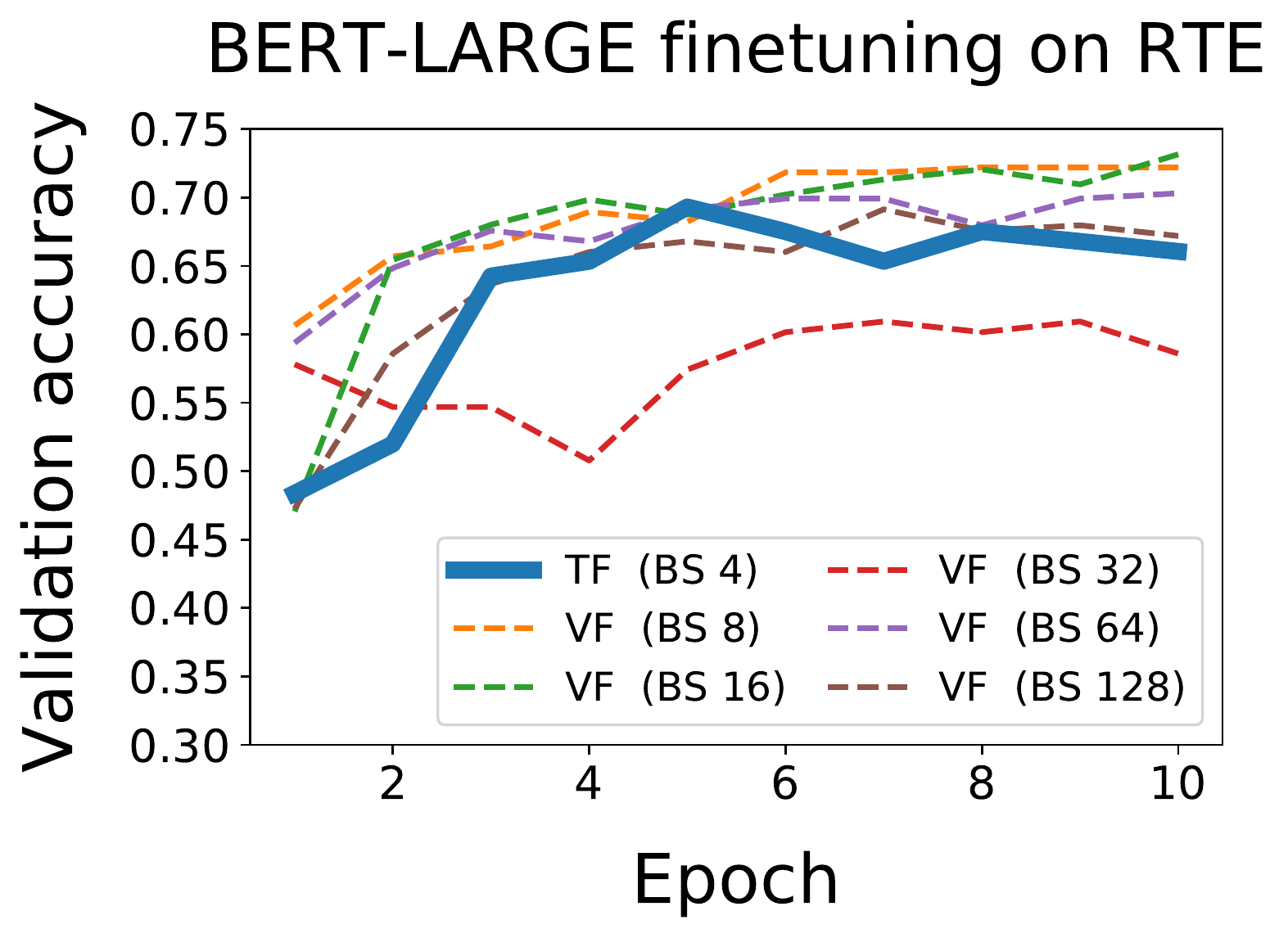}}
  \hspace{1mm}
  \subfigure{\includegraphics[width=0.26\textwidth]{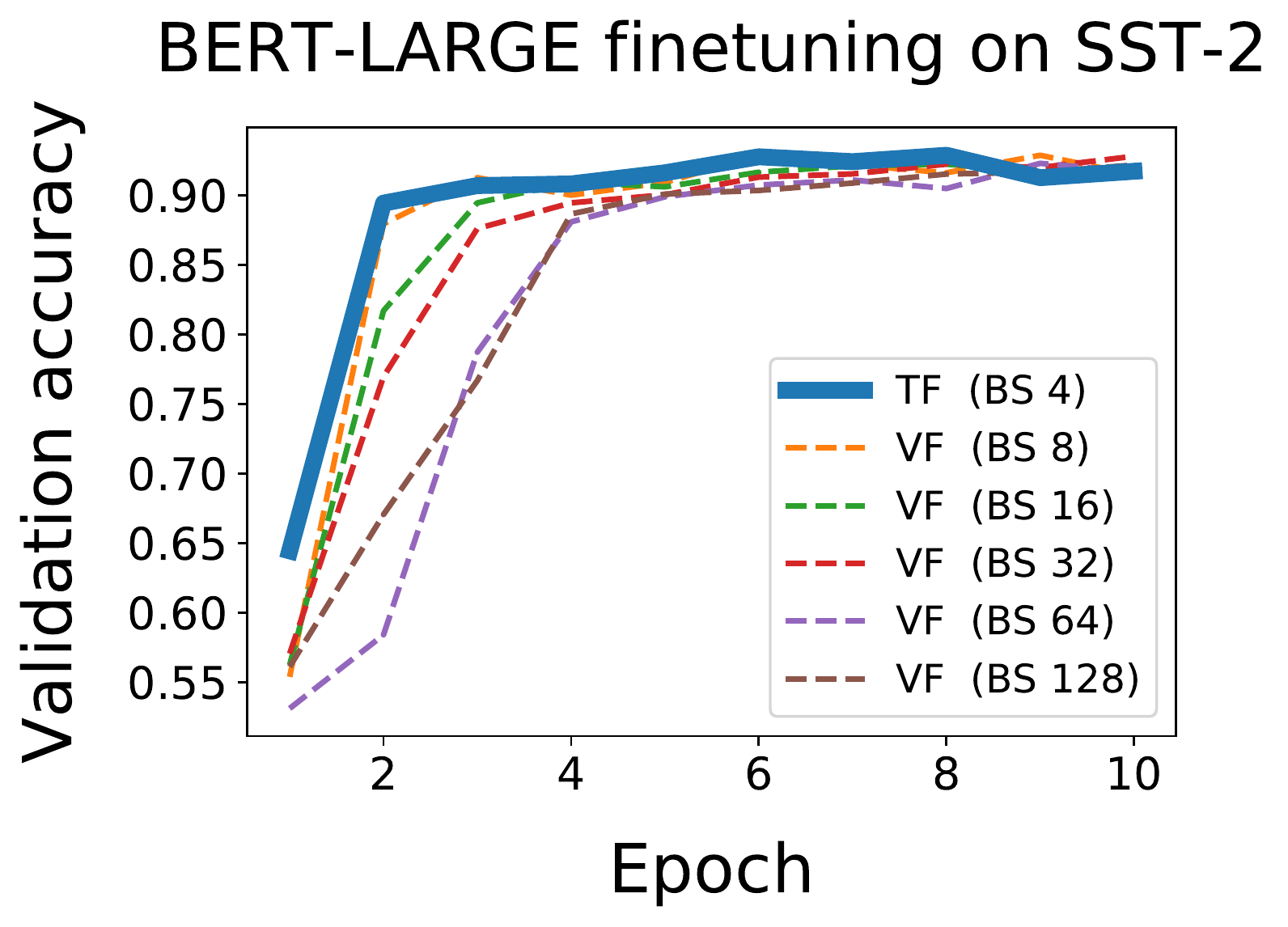}}
  \hspace{1mm}
  \subfigure{\includegraphics[width=0.26\textwidth]{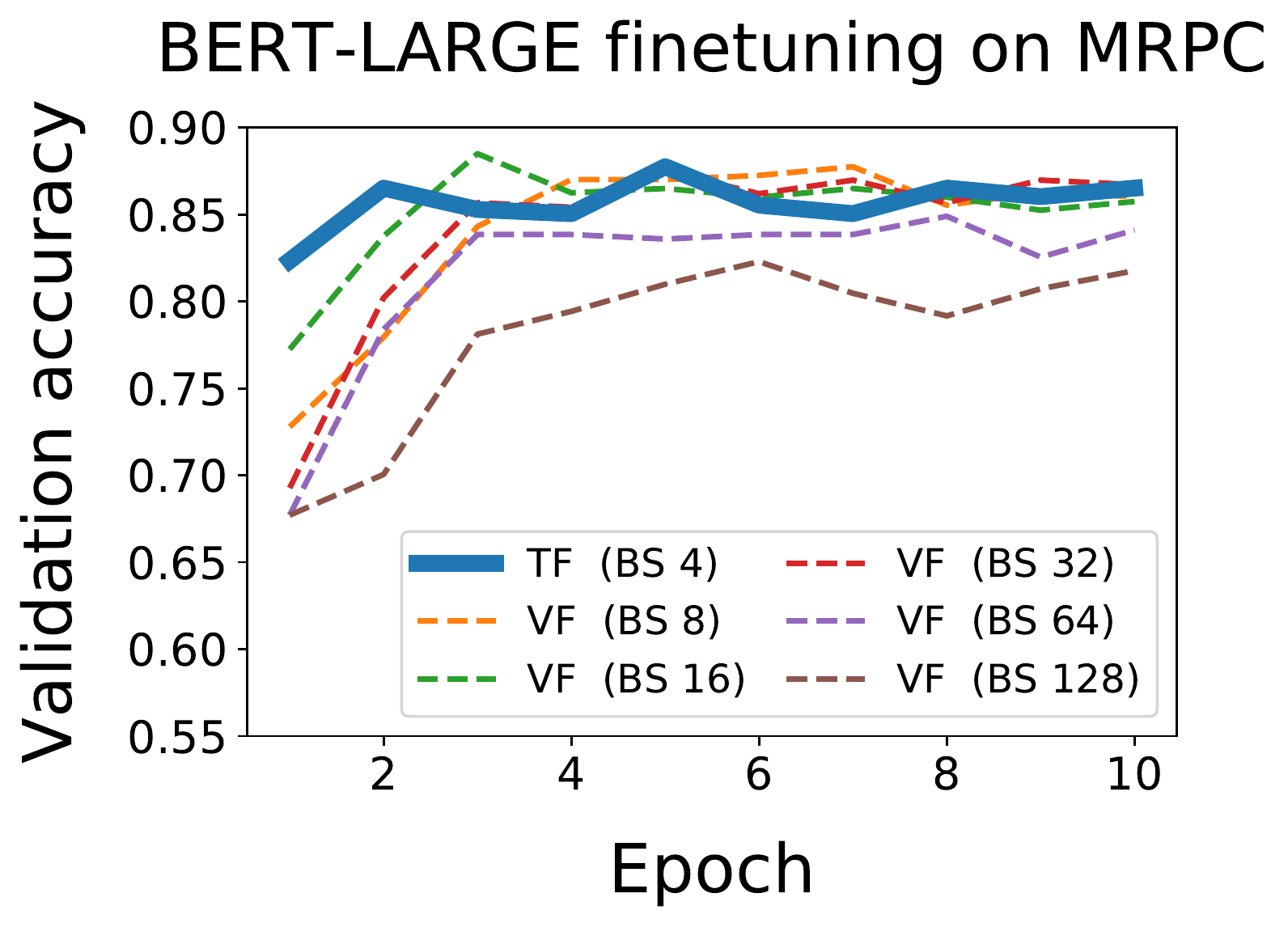}}
  \vspace{-1mm}
  \captionsetup{width=0.92\textwidth}
  \caption{\textbf{Batch size exploration} with \name on a single RTX 2080
    Ti GPU. \name expands the space of possible batch sizes on this
    GPU from 4 (TF) to [4, 8, 16, 32, 64, 128], and can support even
    larger batch sizes. In some cases, such as in RTE (left), being
    able to access larger batch sizes can lead to significantly
    higher final accuracies (+7.1\% with a batch size of 16).}
  \label{fig:bert-large-2080Ti}
\end{figure*}

\subsubsection{ResNet-50 on ImageNet}

In this experiment, we train ResNet-50 on the ImageNet dataset for 90
epochs using a fixed batch size of 8192, a widely used benchmark that
is known to converge to the vicinity of 76\%~\cite{facebook-imagenet, resnet, pytorch-imagenet}.
To demonstrate \name can preserve convergence across GPU types,
we ran this workload on both V100 and RTX 2080Ti GPUs. Each V100
GPU can fit a batch of 256 examples at a given time, so we use 32
total virtual nodes for these runs. For the smaller RTX 2080Ti GPUs,
we use 64 total virtual nodes instead.

Table~\ref{tab:resnet-imagenet-V100} demonstrates \name can reproduce
the target accuracy for all runs ($\pm 0.5\%$) across different numbers
and types of GPUs. Previously, this workload required 32 V100 GPUs.
With \name, however, the user can reproduce the results for the same
workload on even a single GPU. In contrast, attempts to reproduce this
workload on fewer GPUs without retuning the hyperparameters (TF*) led 
to diverged models, \eg doing so on 1 GPU led to a final accuracy of
only 69.25\%, far short of the target 76\%. Additionally, \name preserves
not only the final accuracy but also the entire convergence trajectory
(Figure~\ref{fig:resnet-imagenet-V100}).


\subsubsection{BERT-BASE Finetuning on GLUE}

To demonstrate \name's reproducibility, we also fine-tuned BERT-BASE
on the GLUE dataset using a fixed batch size of 64. The GLUE tasks
considered in this experiment are QNLI, SST-2, and CoLA. For QNLI
and SST-2, we use 1/10th of the original dataset in each epoch and
train for 20 epochs. For CoLA, we train on the whole dataset for
50 epochs.

As with the ResNet workload, \name was able to reproduce the target
accuracies (obtained by running without virtual nodes) across different
numbers of GPUs for all GLUE tasks by preserving the batch size and
the total number of virtual nodes (Table~\ref{tab:bert-base-V100}).
Unlike in the ResNet case, however, the naive approach of not retuning
hyperparameters across different hardware also happened to converge
to the same accuracies in these workloads (not shown). This illustrates
that these workloads are less sensitive to a changing batch size within
this range (8 to 64). While \name did not lead to higher accuracies in
this case, it still guaranteed that results for the batch size of 64
can be consistently reproduced across different sets of resources.

\begin{figure}[t]
  \setlength\columnsep{6pt}
  \vspace{-2.5mm}
  \begin{multicols}{2}
  \centering
  \subfigure[VF scheduler with elasticity]{\includegraphics[width=0.23\textwidth]{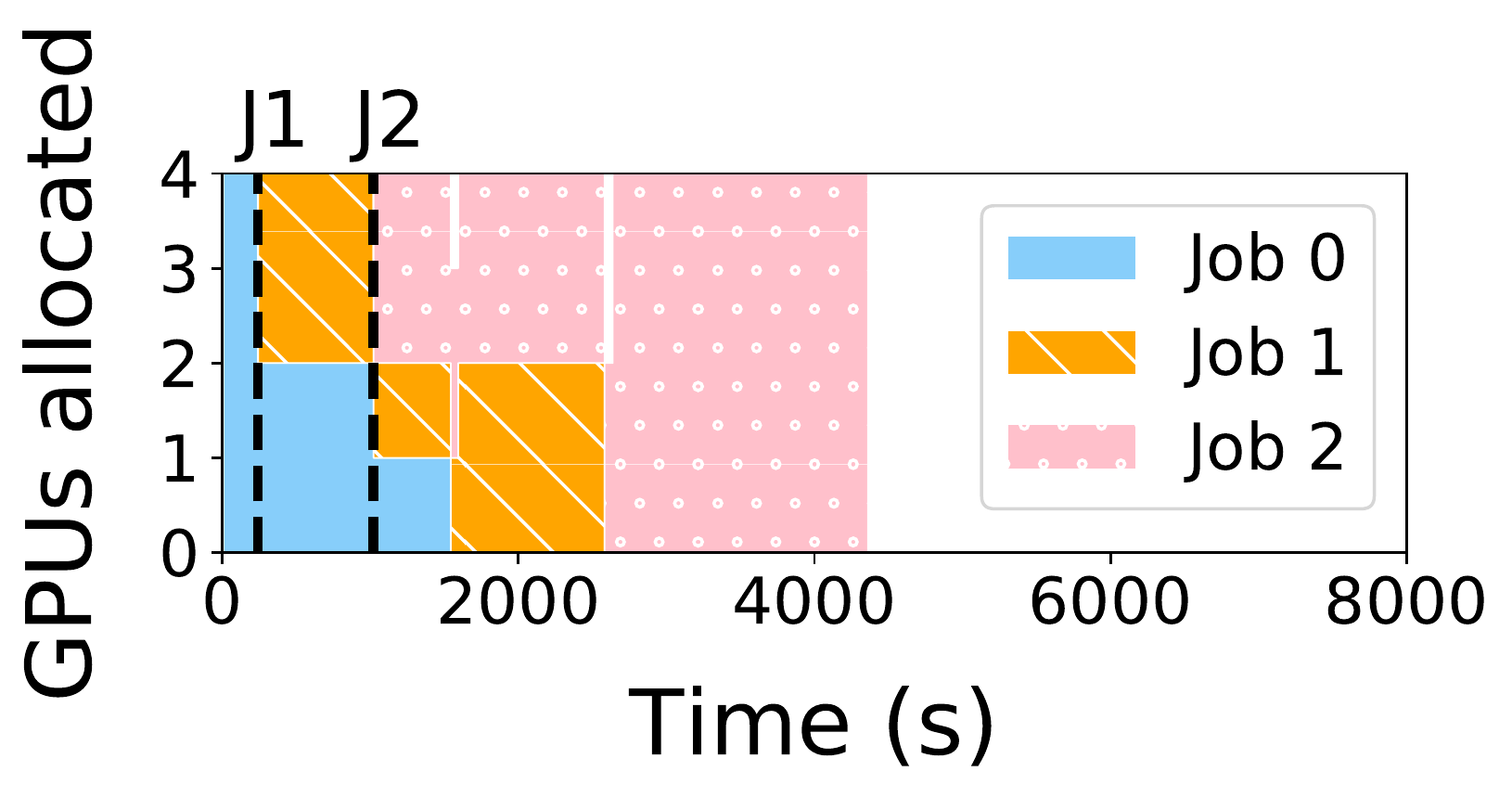}}\vspace{-2mm}
  \subfigure[Priority scheduler, no elasticity]{\includegraphics[width=0.23\textwidth]{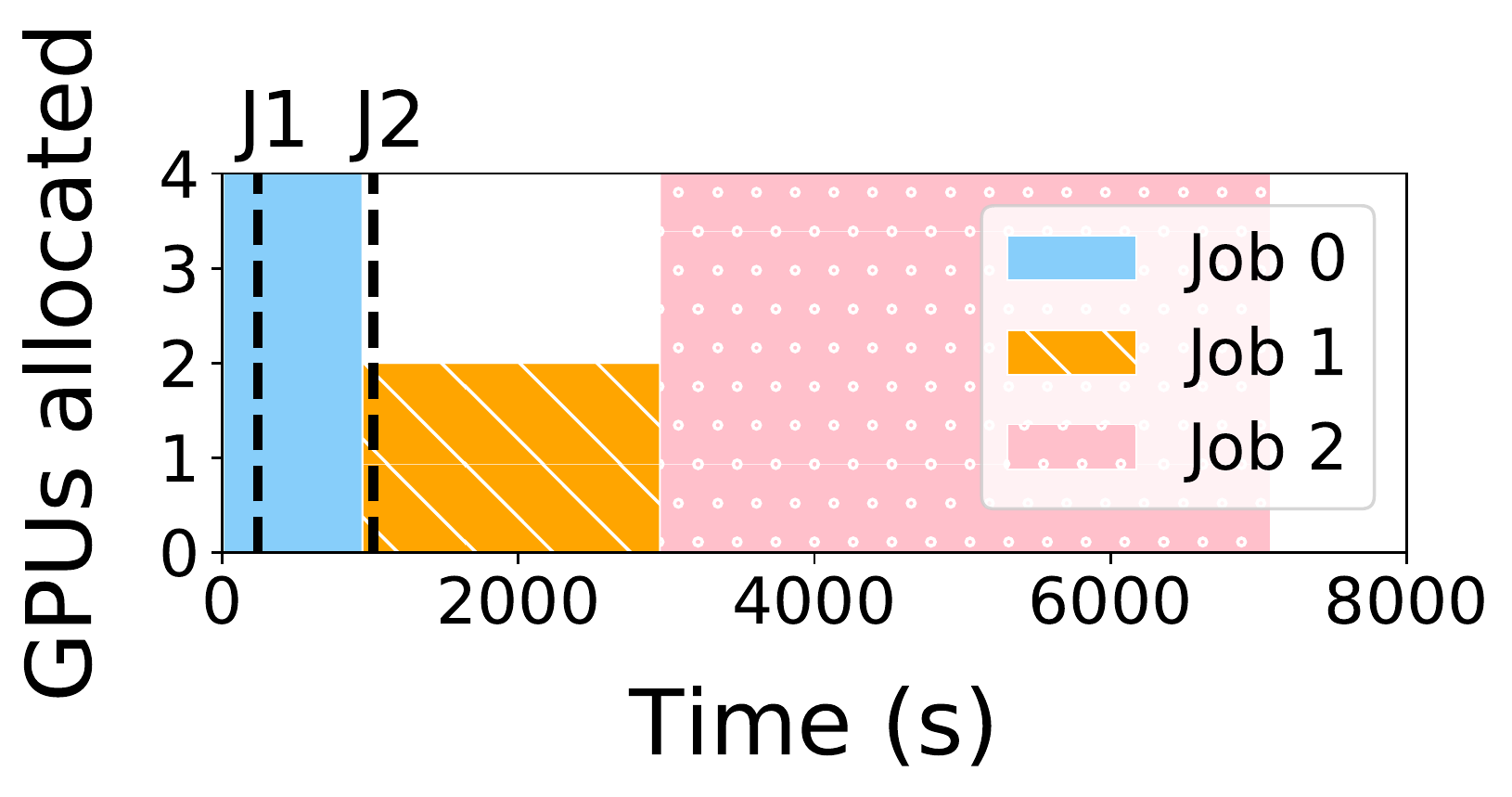}}
  \vspace{-1mm}
  \subfigure[Final accuracy]{\includegraphics[width=0.23\textwidth]{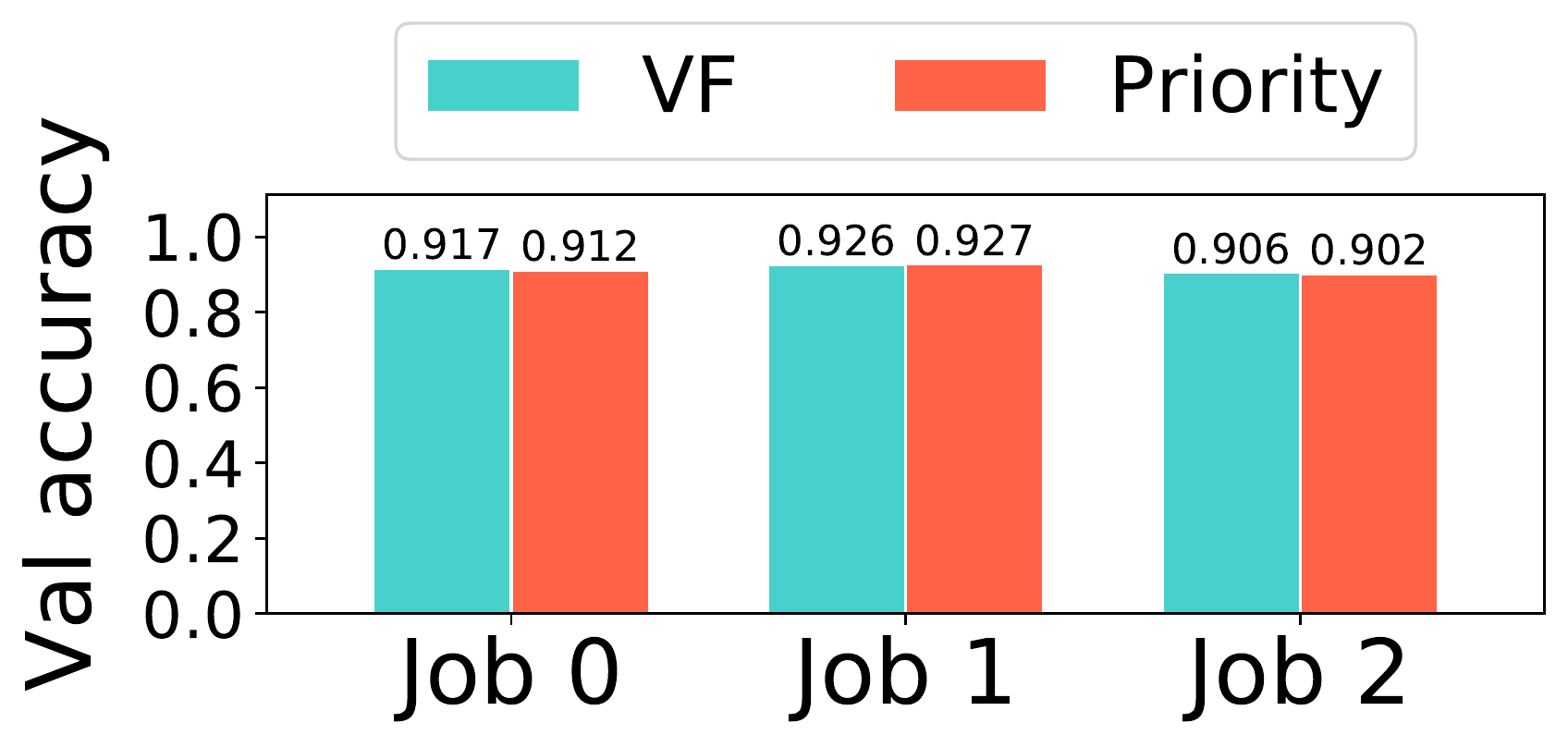}}
  \subfigure[JCT]{\includegraphics[width=0.23\textwidth]{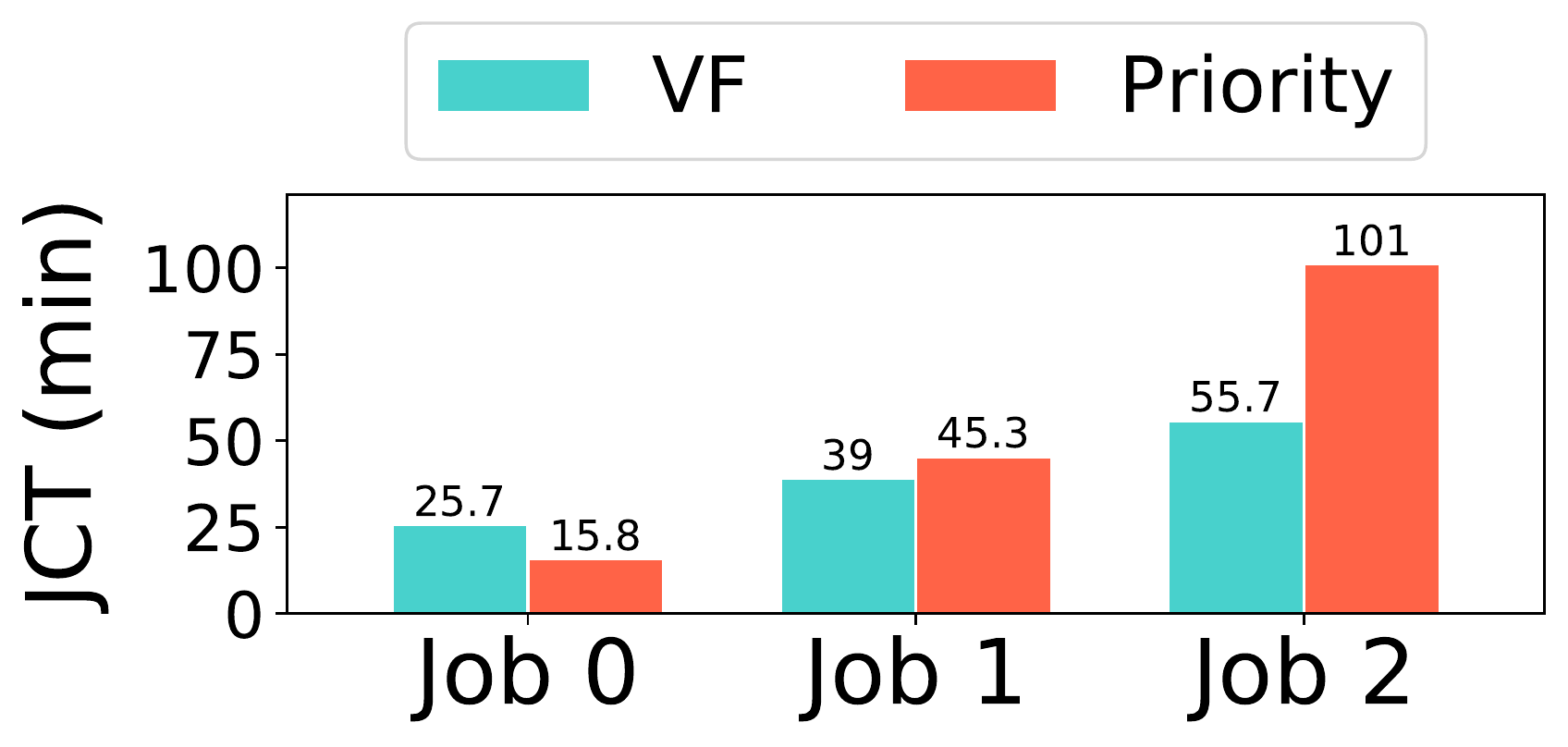}}
  \end{multicols}
  \vspace{-3mm}
  \caption{\textbf{Elasticity} with \name reduces the makespan
    by 38\% and the job completion time (JCT) for the highest priority job by 45\%, while
    preserving model accuracies. In this workload, 3 jobs share 4 V100
    GPUs on a single machine.}
  \label{fig:elasticity-3jobs}
\end{figure}

\subsection{Hyperparameter Exploration}
\label{subsec:exploration}

Another use case of virtual nodes is to explore hyperparameters
previously inaccessible on the same set of resources. To achieve
this, we vary the number of virtual nodes, and consequently the
batch size, while holding the number of GPUs constant, \ie the
opposite of the previous experiments. We fine-tune
BERT-LARGE on three GLUE tasks, RTE, SST-2, and MRPC, for 10
epochs on a single RTX 2080Ti GPU.

Figure~\ref{fig:bert-large-2080Ti} plots the model convergence
for this experiment. Unlike before, since the batch size changes
across runs, so do the convergence trajectory and the final
accuracy. This allows the user to explore the convergence
characteristics of various batch sizes, without deploying the
resources that would have been necessary to run these batch
sizes using vanilla TensorFlow (\eg 32 GPUs for a batch size
of 128).

In some cases, \name can even achieve higher accuracies in the
batch sizes explored. For RTE (a reading entailment task), using
a larger batch size of 16 is now possible on 1 GPU, even though
this batch size required 4 GPUs before. This configuration ended
up improving the final accuracy by 7 percentage points on the
same set of resources.


\begin{table}[t]
  \small
  \centering
  \setlength{\abovecaptionskip}{4mm}
  \begin{tabular}{ccccc}
    \noalign{\hrule height 1.5pt}
    Model & Dataset & Batch sizes & VN$_{GPU}$ \\
    \noalign{\hrule height 1pt}
    ResNet-56 & cifar10 & 64, 128 & 1 \\
    [0.1cm]ResNet-50 & ImageNet &
      \pbox{20cm}{\relax\ifvmode\centering\fi 256, 512, 1024\\2048, 4096, 8192} & 1, 2, 4\\[0.25cm]
    BERT-BASE & CoLA & 8, 16, 32, 64, 128 & 1, 2\\
    BERT-BASE & SST-2 & 8, 16, 32, 64, 128 & 1, 2\\
    [0.1cm]Transformer & WMT & \pbox{20cm}{\relax\ifvmode\centering\fi 4096, 8192, 16384\\32768, 65536} & 1, 2\\[0.25cm]
    \noalign{\hrule height 1.5pt}
  \end{tabular}
  \caption{\textbf{Elasticity:} Mix of workloads used in 20 job experiment.
    Each job in the trace is selected uniformly at random from this
    set of workloads and assigned a random priority chosen from
    (1, 5, 10).}
  \label{tab:elasticity-20jobs-workload}
\end{table}

\subsection{Elasticity}
\label{subsec:elasticity}

Another important use case enabled by \name is resource elasticity:
a job can be resized dynamically during training by adjusting the
number of virtual nodes per GPU. This section describes experimental
results that highlight the cluster-level benefits of this approach.

\subsubsection{Elastic Scheduling with Three Jobs}

Using the scheduling framework described in \S\ref{subsec:wfs-scheduler},
we ran two traces with and without \name. The first is a
simple 3-job trace designed to illustrate a scenario in which
elasticity can have significant effects on cluster-level
objectives. Job~0 fine-tunes BERT-BASE on SST-2, Job~1 trains
ResNet-56 on cifar10~\cite{cifar10}, and Job~2 fine-tunes BERT-BASE on QNLI.
The BERT jobs both demand 4 GPUs, while the ResNet job demands
only 2 GPUs. The jobs arrive in the order of increasing priority,
with Job 2 being the highest.

Figure~\ref{fig:elasticity-3jobs} compares running this trace
with the \name scheduler, which dynamically resizes jobs to satisfy
cluster-level Weighted Fair Shares (WFS), to running it with a simple
priority scheduler that orders jobs in descending priority but
does not resize any job. With \name, existing jobs downsize
as soon as a new job with priority arrives. With the static
priority scheduler, however, the high priority Job~2 is stuck
behind Job~1 for a long time, leaving 2 GPUs idle for the
entire duration of Job~1.

Observe that although all 3 jobs resized over the course of
their respective lifetimes in the \name case, they all converged
to the same accuracies as their counterparts in the simple
priority scheduler case. Thus, the \name scheduler is able to
reduce the makespan by 38\% and the high priority job completion time (JCT) by 45\%,
while preserving the application-level semantics of each job.

\begin{figure}[t]
  \centering
  \vspace{-2mm}
  \setlength\belowcaptionskip{-5mm}
  \subfigure{\includegraphics[width=0.35\textwidth]{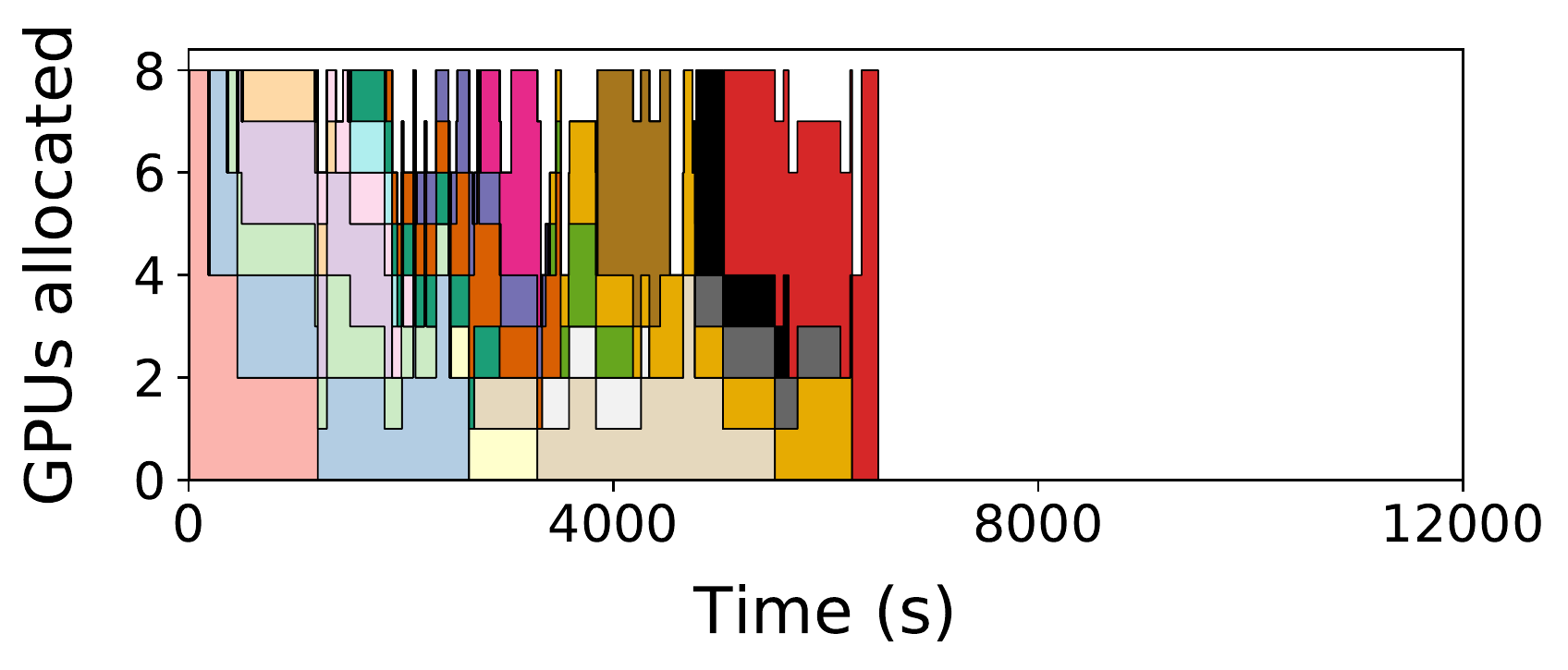}}\\[-1mm]
  \vspace{-8mm}
  \subfigure{\includegraphics[width=0.35\textwidth]{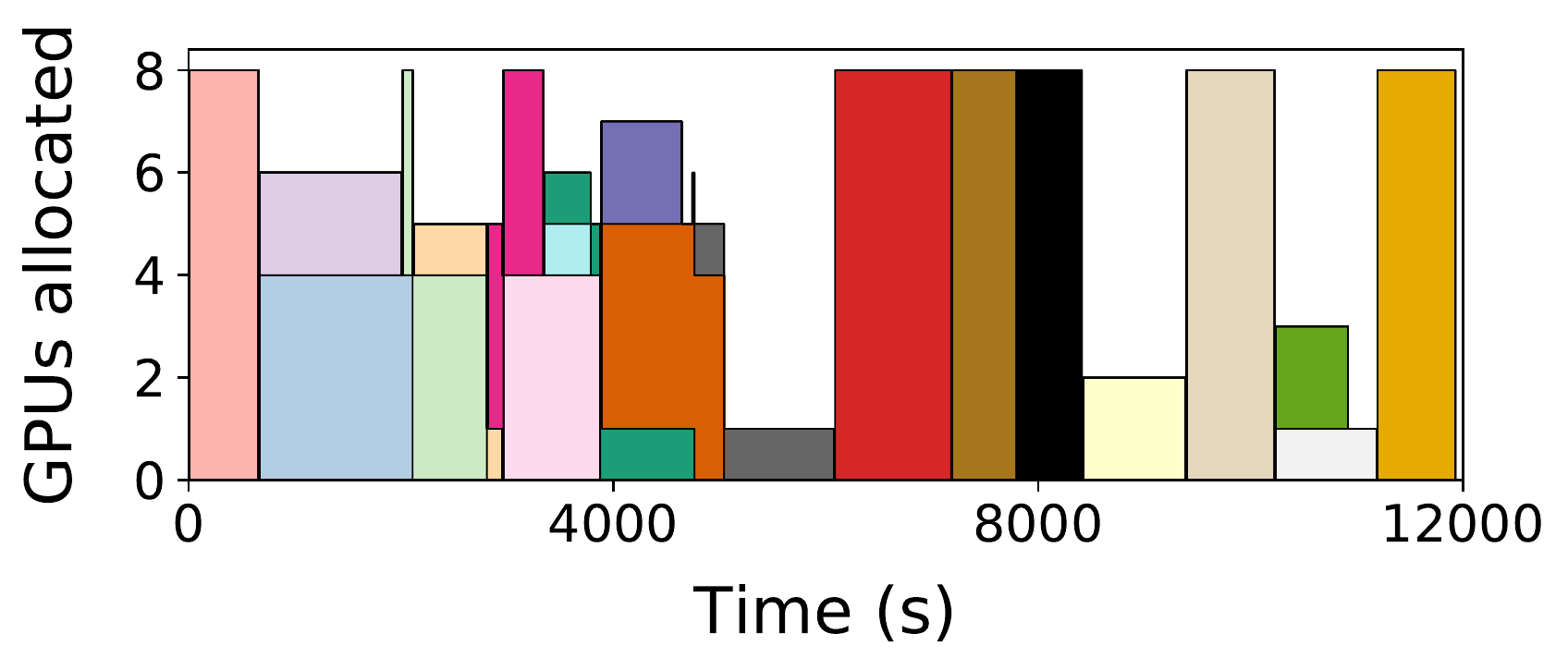}}
  \vspace{-3mm}
  \caption{\textbf{Elasticity} with \name (top) increases average cluster
    utilization by 19.5\% and reduces makespan by 45.5\%, compared to a
    simple priority scheduler (bottom) that does not perform elasticity.
    Each colored box corresponds to a job. Boxes resize for the elastic
    scheduler (top) but not for the static scheduler (bottom).}
  \label{fig:elasticity-20jobs}
\end{figure}

\begin{figure}[t]
  \centering
  \subfigure{\includegraphics[width=0.2\textwidth]{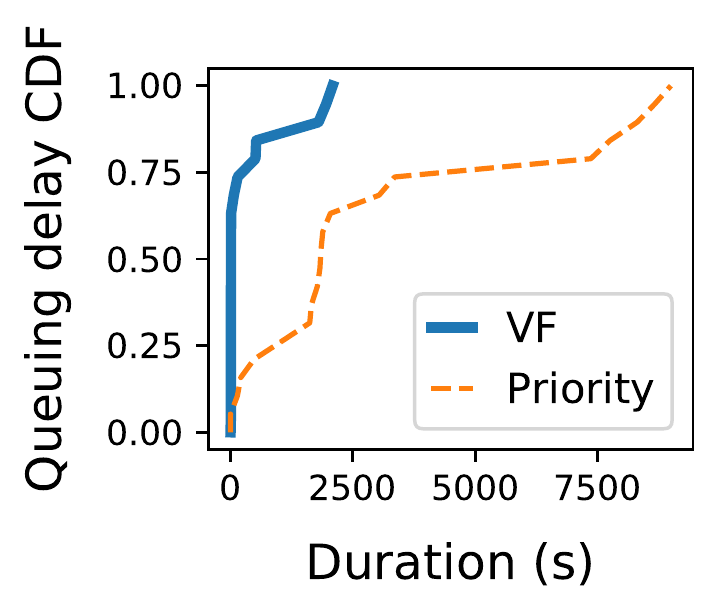}}
  \subfigure{\includegraphics[width=0.2\textwidth]{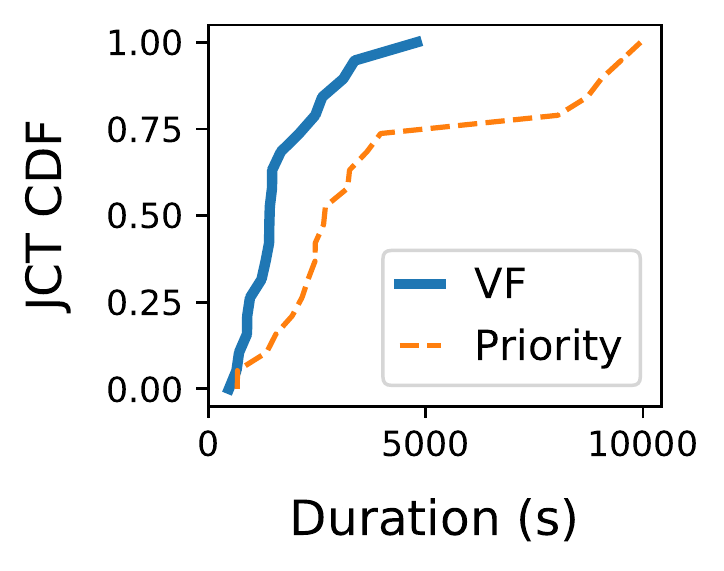}}
  \vspace{-3mm}
  \caption{\textbf{Elasticity:} In the same 20 job experiment shown in
    Figure~\ref{fig:elasticity-20jobs}, \name reduces the
    median JCT by 47.6\% and the median queuing delay by 99.3\%
    by resizing jobs dynamically.}
  \label{fig:elasticity-20jobs-cdf}
\end{figure}

\begin{table}[t]
  \small
  \centering
  \setlength\dashlinedash{0.75pt}
  \setlength\dashlinegap{2pt}
  \setlength\arrayrulewidth{0.75pt}
  \setlength{\tabcolsep}{4pt}
  \setlength{\belowcaptionskip}{-3mm}
  \begin{tabular}{r|ccc|ccc}
    \noalign{\hrule height 1.5pt}
    & \multicolumn{3}{c|}{V100} & \multicolumn{3}{c}{P100}\\
    Exp & Num & BS$_{GPU}$ & VN$_{GPU}$ & Num & BS$_{GPU}$ & VN$_{GPU}$\\
    \noalign{\hrule height 1pt}
    H1a & 1+1 & 2048 & 8 & 1+1 & 2048 & 8 \\
    b & 1+1 & 3072 & 16 & 1+1 & 1024 & 4 \\
    c & 1+1 & 3072 & 32 & 1+1 & 1024 & 4 \\
    \hline
    H2a & 1+1 & 3072 & 16 & 2+2 & 512 & 2 \\
    b & 1+1 & 3072 & 16 & 2+2 & 512 & 4 \\
    c & 1+1 & 3072 & 16 & 2+2 & 512 & 8 \\
    d & 1+1 & 3072 & 16 & 2+2 & 512 & 16 \\
    \hline
    H3 & 1+1 & 2048 & 8 & 4+4 & 512 & 2 \\
    \noalign{\hrule height 1.5pt}
  \end{tabular}
  \vspace{2mm}
  \caption{\textbf{Heterogeneous training} configurations for ResNet-50
    on ImageNet (batch size 8192). Columns BS$_{GPU}$ and VN$_{GPU}$
    refer to the batch size and number of virtual nodes assigned to each
    GPU of the given type respectively, and 1+1 in the Num column refers
    to 2 servers with 1 GPU each.}
  \label{tab:het-configs}
\end{table}

\begin{figure}[t]
  \small
  \setlength{\abovecaptionskip}{6mm}
  \setlength{\belowcaptionskip}{2mm}
  \setlength{\tabcolsep}{2pt}
  \includegraphics[width=0.72\linewidth, trim=0 1.75cm 0 0]{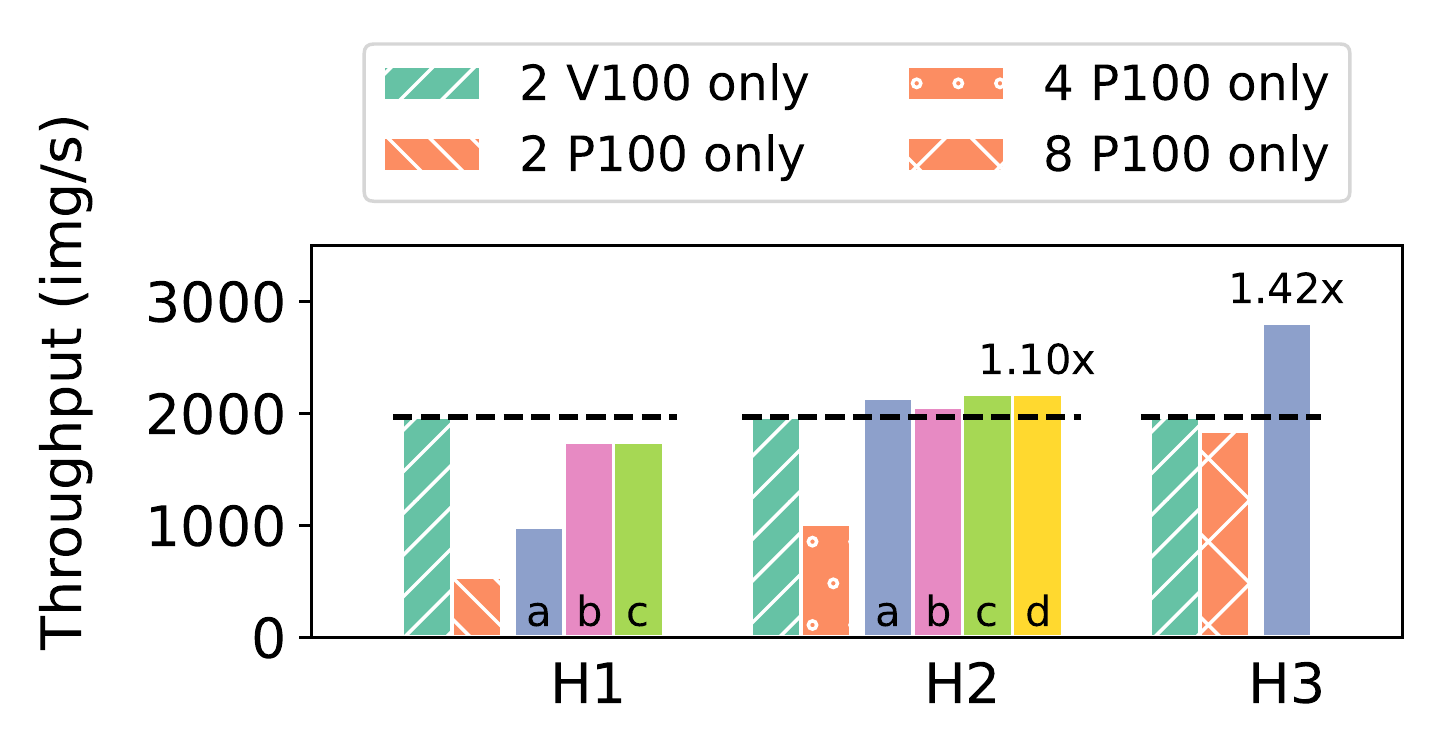}
  \hspace{1mm}
  \begin{tabular}[b]{cc}\noalign{\hrule height 1.5pt}
    & Acc (\%) \\\noalign{\hrule height 1pt}
    H1b & 75.97 \\
    H2a & 75.92 \\
    H3 & 75.80 \\
    \noalign{\hrule height 1.5pt}
  \end{tabular}
  \caption{\textbf{Heterogeneous training} can improve throughput by up
    to 42\% while converging to same target accuracy
    (76\%~\cite{facebook-imagenet}) as homogeneous training. Experiment
    H3 had the largest improvement because the V100 only performance
    is most similar to the P100 only performance. The specific
    configurations can be found in table~\ref{tab:het-configs}.}
  \label{fig:het-throughputs}
\end{figure}

\subsubsection{Elastic Scheduling with Twenty Jobs}

Next, we evaluate \name on a more realistic trace consisting of 20
jobs arriving according to a poisson distribution, with an average
load of 12 jobs per hour (average interarrival time of 5 minutes).
The mixture of workloads used in this trace is selected uniformly
at random from Table~\ref{tab:elasticity-20jobs-workload}. To speed
up the experiment, we train each job for only a subset of the steps
or epochs needed for convergence.

Figure~\ref{fig:elasticity-20jobs} depicts the GPU allocations
for both schedulers over time. Compared to the simple priority
scheduler, enabling elasticity with \name
improved average cluster utilization from 71.1\% to 90.6\%,
reduced the makespan by 45.5\%, the median JCT by 47.6\%,
and the median queuing delay by 99.3\%. The largest gain
from using elasticity is the reduction in queuing delay
(Figure~\ref{fig:elasticity-20jobs-cdf}): most jobs are
assigned some GPUs as soon as they are submitted instead
of being queued behind other potentially long jobs. This is
especially true for high priority jobs, which can partially
preempt lower priority jobs by downsizing them.

\subsection{Heterogeneous Training}
\label{subsec:het}

In this section, we explore the conditions under which heterogeneous
training in \name is most beneficial and evaluate the effectiveness of
the heterogeneous solver~(\S\ref{subsec:het-assignment})
in searching for an efficient configuration.

\begin{figure}[t]
  \begin{subfigure}{}
    \centering
    \setlength{\abovecaptionskip}{0mm}
    \setlength{\belowcaptionskip}{-2mm}
    \includegraphics[width=0.8\linewidth]{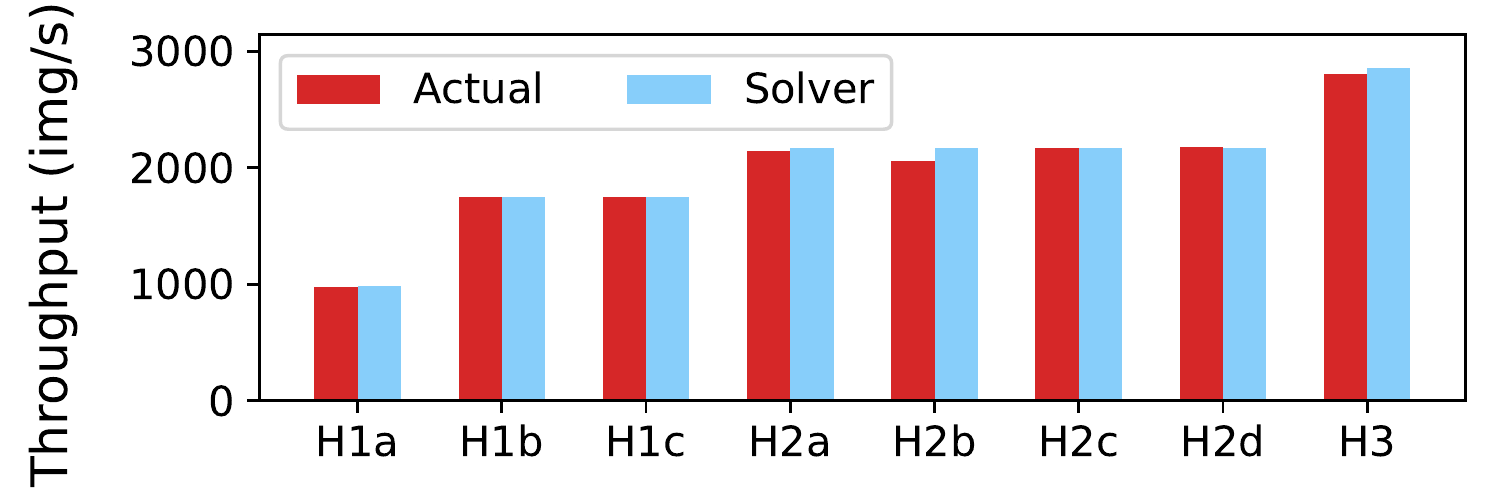}
    \caption{\textbf{Heterogeneous solver} produces throughputs within 5.6\%
      of actual throughputs on average (experiments from Table~\ref{tab:het-configs}).}
    \vspace{3mm}
    \label{fig:het-solver-throughputs}
  \end{subfigure}{}
  \begin{subfigure}{}
    \centering
    \setlength{\abovecaptionskip}{0mm}
    \setlength{\belowcaptionskip}{-2mm}
    \includegraphics[width=0.8\linewidth]{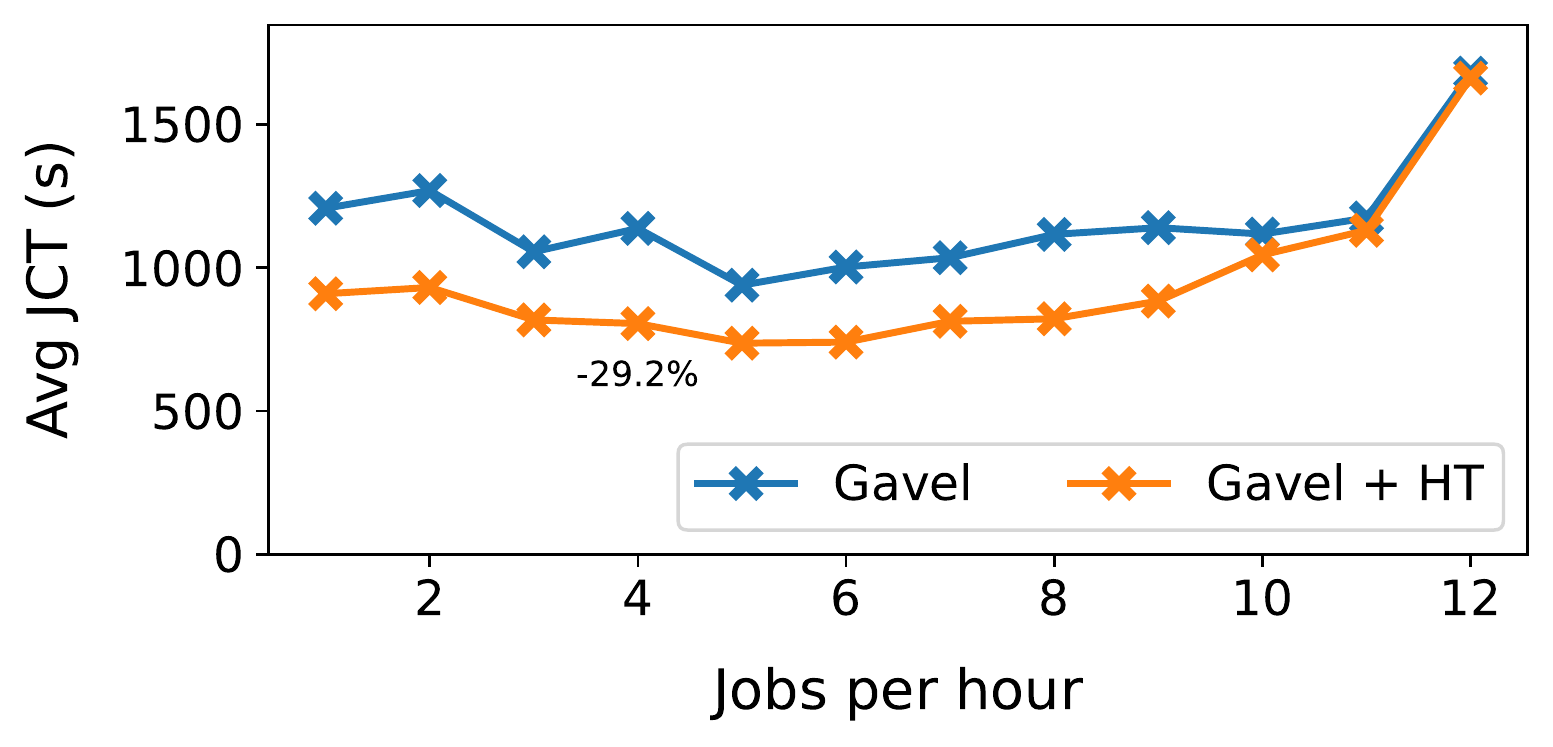}
    \caption{\textbf{Heterogeneous scheduler:} Extending Gavel~\cite{gavel}
      to additionally consider heterogeneous allocations can reduce the average
      job completion time by up to 29.2\%. The cluster consists of 4 V100 GPUs,
      8 P100 GPUs, and 16 K80 GPUs. (Simulation)}
    \vspace{3mm}
    \label{fig:het-scheduler}
  \end{subfigure}{}
  \begin{subfigure}{}
    \centering
    \setlength{\abovecaptionskip}{0.5mm}
    \setlength{\belowcaptionskip}{1mm}
    \includegraphics[width=0.8\linewidth]{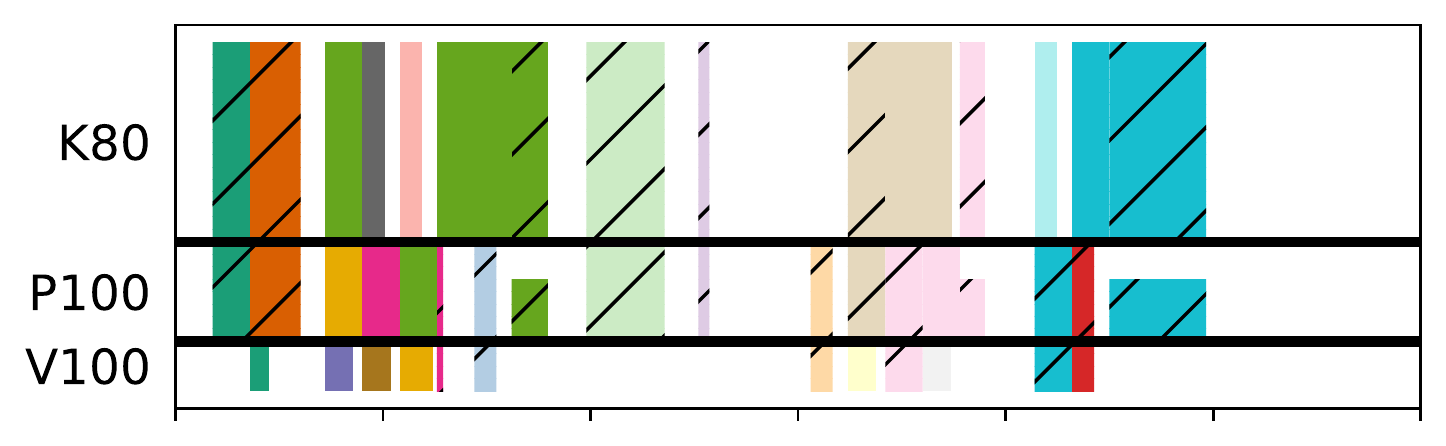}
    \includegraphics[width=0.8\linewidth]{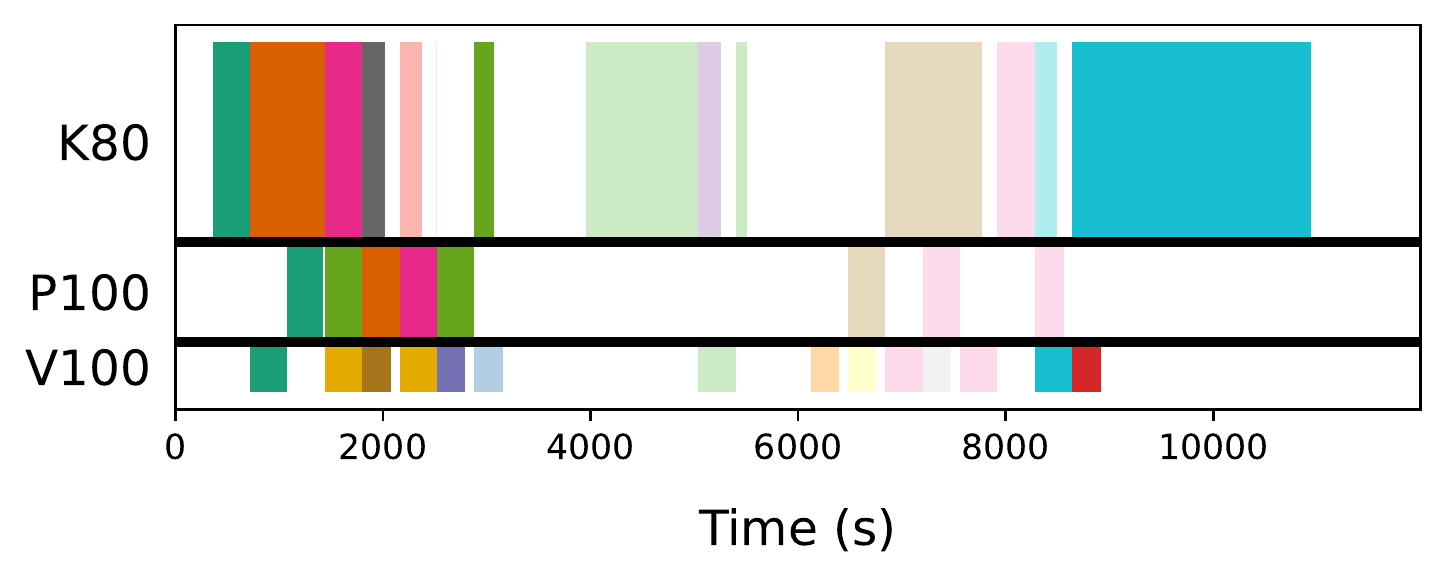}
    \caption{\textbf{Heterogeneous scheduler:} An example trace
      where Gavel~\cite{gavel} with heterogeneous allocations (top)
      reduces the average job completion time by 26.4\% compared to
      Gavel without (bottom). Each colored box refers to an allocation,
      and each box with a hatched pattern refers to a heterogeneous
      allocation. In this trace, 8 jobs arrive per hour on average.
      (Simulation)}
    \label{fig:het-scheduler-8jph}
  \end{subfigure}
\end{figure}

\subsubsection{Throughput and Accuracy}
\label{subsubsec:het-throughput}

Figure~\ref{fig:het-throughputs} demonstrates the effectiveness of
heterogeneous training across different sets of resources compared
to homogeneous training. Detailed configurations regarding these
experiments can be found in Table~\ref{tab:het-configs}.


Heterogeneous configuration H3 significantly outperformed both
the V100 only (by 42.3\%) and the P100 only (by 52.4\%) homogeneous
configurations. Compared to H1 and H2, H3 is best able to balance
the step times of the two individual GPU types. This is because,
for this workload, V100 GPUs are roughly 4x as fast as P100 GPUs,
and H3 uses 2 V100 GPUs + 8 P100 GPUs. Importantly, this configuration
also converged to the target accuracy of 76\%~\cite{facebook-imagenet,
resnet, pytorch-imagenet}, the same as homogeneous training.

\name's heterogeneous solver accurately predicted throughputs for this
set of experiments (Figure~\ref{fig:het-solver-throughputs}).
In the H1 group, the V100 only configuration is the most efficient
because there are not enough P100 GPUs to compensate for the difference
in performance. For this group, \name's heterogeneous solver fell back
on recommending the more efficient V100 only configuration. 

\subsubsection{Heterogeneous Scheduler}
\label{subsubsec:het-scheduler}

To illustrate the benefits of heterogeneous training in
a multi-tenant cluster, we extended Gavel~\cite{gavel} to additionally
consider heterogeneous allocations. Although Gavel was designed for
heterogeneous clusters, it only considers \textit{homogeneous}
allocations. We evaluate our implementation in a simulated environment
consisting of 4 V100, 8 P100, and 16 K80 GPUs, drawing from a subset
of the workloads in Table~\ref{tab:elasticity-20jobs-workload}.
Following~\cite{gavel}, we use a round duration of 6 minutes and
their formulation of the Least Attained Service (LAS) objective.

In this experiment, using heterogeneous allocations allowed the
scheduler to reduce the average job completion time by up to 29.2\%
(Figure~\ref{fig:het-scheduler}). At higher job arrival rates, the
benefits of using heterogeneous allocations diminishes, however, and
the system gracefully falls back to prior behavior. This is because
leftover resources can be allocated to new jobs instead.
Figure~\ref{fig:het-scheduler-8jph} illustrates an example of how
heterogeneous training can improve job completion time. Individual
jobs can train faster on multiple types of GPUs if there are idle
resources, \eg the rightmost job's throughput improved by 33.7\% with
5 extra P100 GPUs in addition to the 16 K80 GPUs already assigned to it.

\subsection{Microbenchmarks}
\label{subsec:microbenchmarks}

\begin{figure}[t]
  \centering
  \vspace{-1mm}
  \setlength\belowcaptionskip{-5mm}
  \begin{subfigure}{}
    \includegraphics[width=0.4\textwidth]{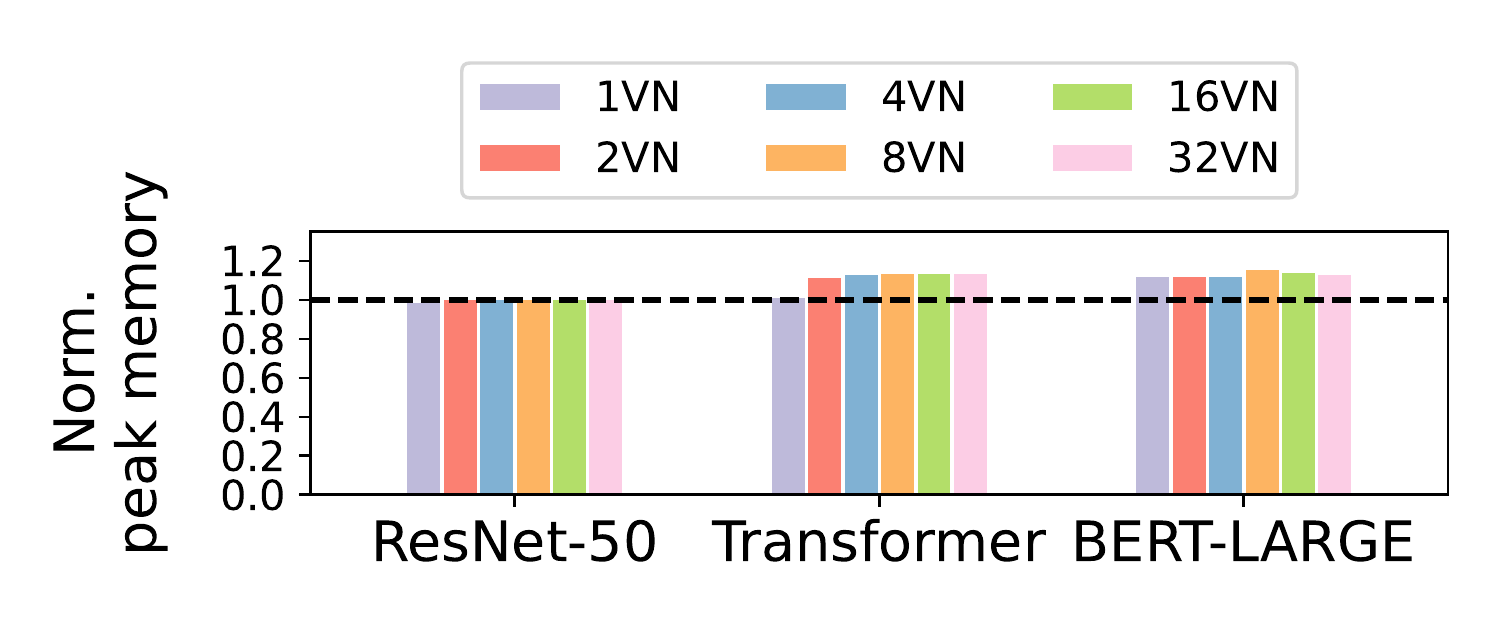}
  \end{subfigure}
  \vspace{-5mm}
  \begin{subfigure}{}
    \hspace*{-1mm}\includegraphics[width=0.4\textwidth]{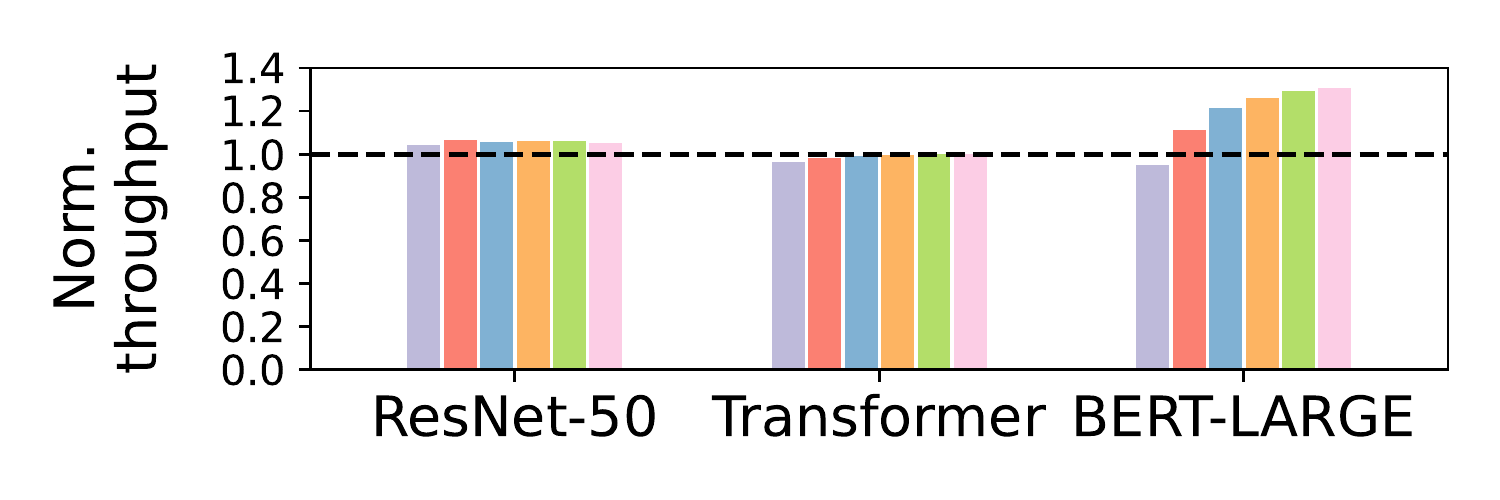}
    \vspace{-1mm}
    \caption{Peak memory and throughput on a single RTX 2080Ti GPU,
      normalized by the values from TensorFlow. Memory overhead
      scales with the model size and is constant across virtual nodes
      (top), while throughput scales with the number of virtual nodes
      for large models, due to fewer model updates (bottom).}
    \label{fig:microbenchmark}
  \end{subfigure}
\end{figure}

\begin{figure}[t]
  \centering
  \setlength{\belowcaptionskip}{0mm}
  \hspace*{-4mm}\includegraphics[width=0.85\linewidth]{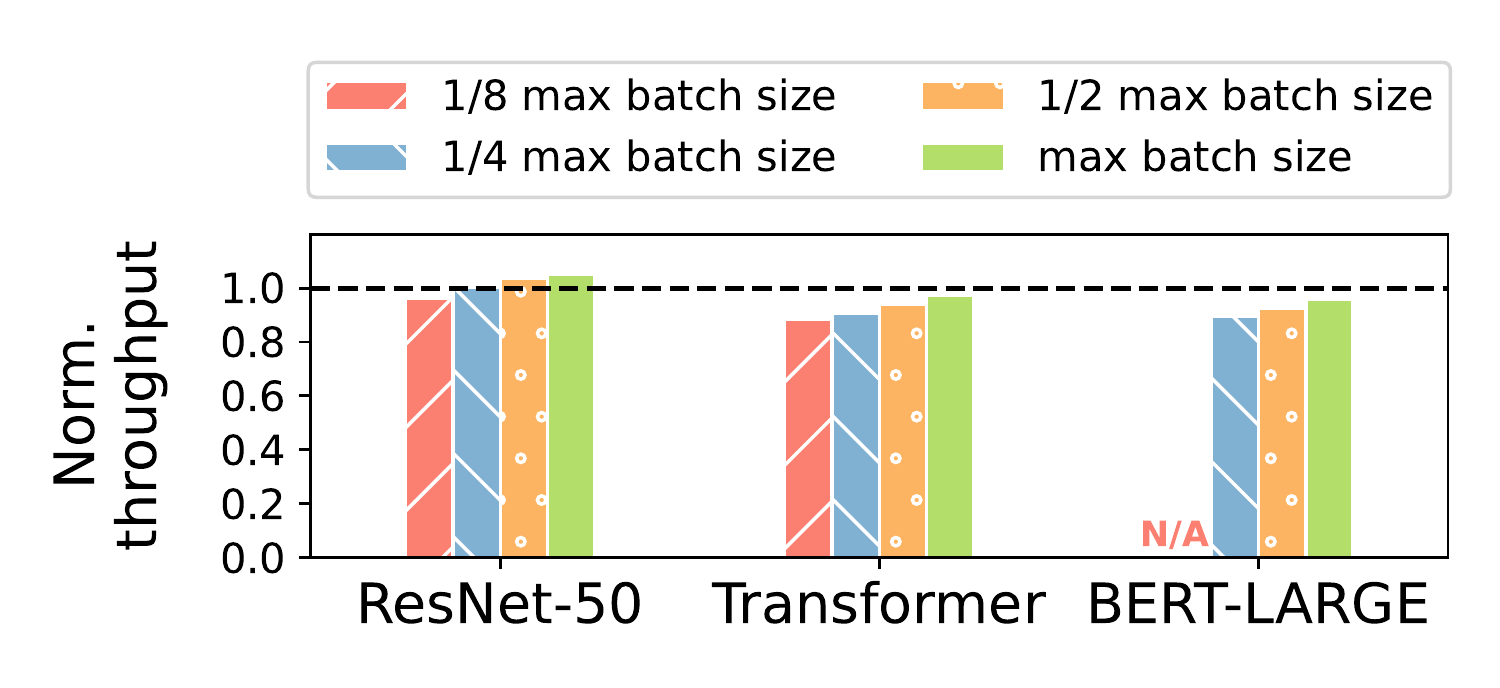}
  \vspace{-2mm}
  \caption{Overhead on a single RTX 2080Ti GPU for batch sizes that
    fit within the GPU's memory, normalized by throughputs from
    TensorFlow. The max batch sizes for ResNet-50, Transformer, and
    BERT-LARGE on this GPU are 192, 3072, and 4 respectively.}
  \label{fig:overhead}
\end{figure}

Virtual node processing adds a gradient buffer to aggregate gradients
across virtual nodes. This gradient buffer is the same size of the model
(Figure~\ref{fig:microbenchmark} top): larger models like BERT see up to
16.2\% memory overhead. Beyond 2 virtual nodes, however, this memory overhead
stays constant.

Figure~\ref{fig:microbenchmark} (bottom) plots the throughput across a
range of virtual nodes for the same three workloads. The global batch
size increases with the number of virtual nodes, and so the frequency
of potentially expensive model updates decreases proportionally.
For these workloads, using virtual nodes at worst lowers the throughput
by 4.2\% but can increase it by 31.4\% in some cases, especially when
the model is large (BERT), since updating large models is expensive.

Figure~\ref{fig:overhead} plots the overhead for workloads that already
fit within the accelerator memory. In all workloads considered, the overhead
is minimal; the throughput of using virtual nodes is within 88.4\% of the
throughput without using virtual nodes. Note that for these single accelerator
workloads, the user can simply disable virtual nodes since the job likely will
not benefit from elasticity or heterogeneous training.

\section{Future Directions}
\label{sec:future}

The virtual node abstraction is not limited to the above use cases.
For instance, \name can be extended to support:

\textbf{Fault tolerance.} We can reuse existing elasticity
mechanisms to migrate virtual nodes from failed workers to remaining
healthy ones, and then to the new replacement workers when they become
available. This would ensure training is uninterrupted from the
application's perspective. In contrast, state-of-the-art solutions
must restart the job from potentially stale checkpoints if even a
single worker fails~\cite{checkfreq}, since the model graph does not
support changes in cluster membership.

\begin{figure}[t]
  \centering
  \includegraphics[width=0.95\linewidth,trim={0.25cm 4cm 3cm 0}]{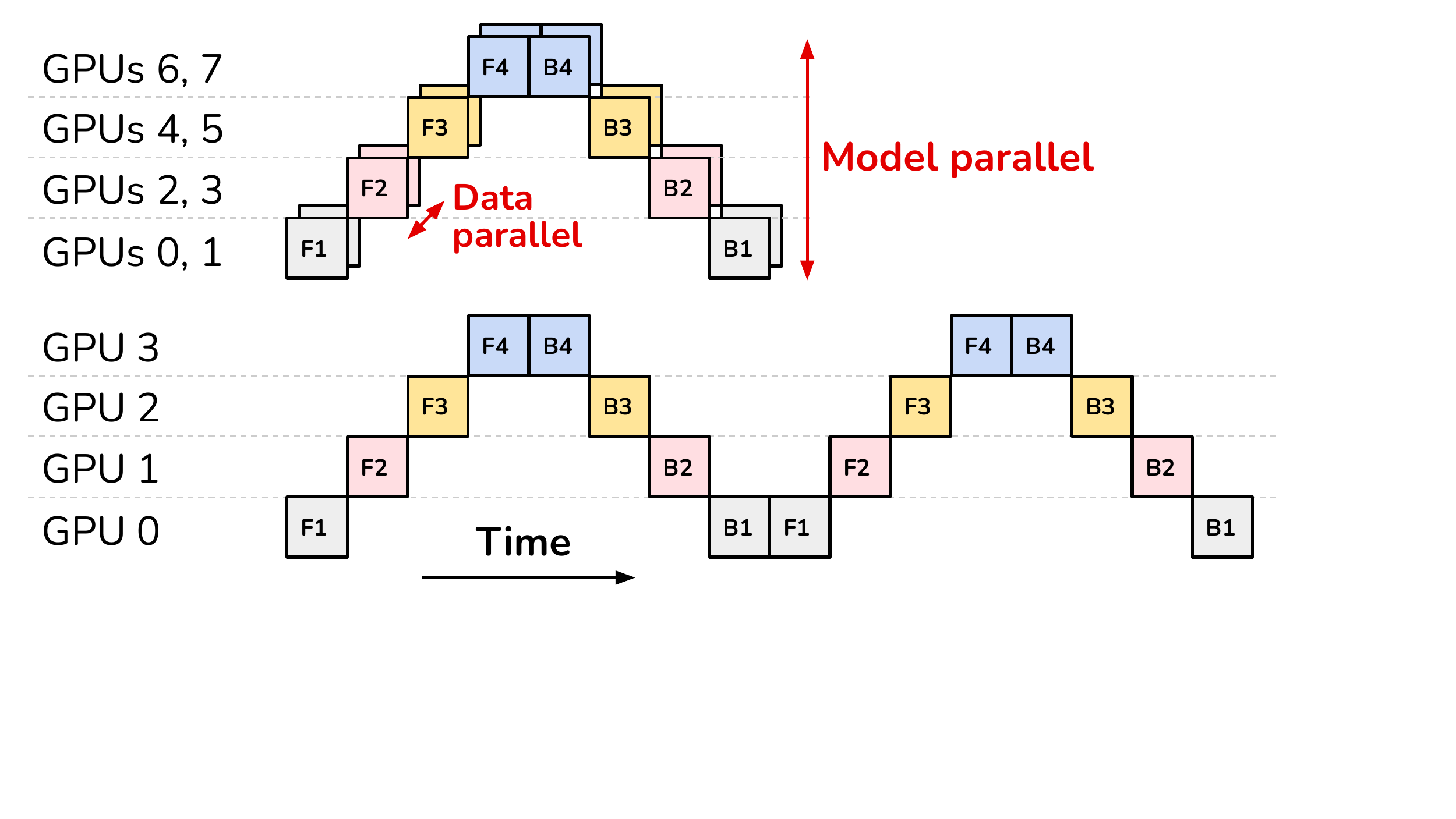}
  \caption{Model parallelism combined with data parallelism today (top),
    versus model parallelism with virtual nodes (bottom), which lowers
    the resource requirement for this workload by half. This can be
    further optimized by pipelining the virtual nodes, which would overlap
    boxes F1 and F2 for example, as in GPipe~\cite{gpipe}.}
  \label{fig:model-parallelism}
\end{figure}

\textbf{Model parallelism.} Training extremely large models~\cite{gpt3,
turing-nlg, megatron-lm} relies on \textit{model parallelism}, which
partitions, instead of replicates, the model across the accelerators
in the system. The model can be partitioned by layer or groups of layers
(as in pipeline parallelism~\cite{gpipe}) and/or by slices within each layer
(spatial partitioning~\cite{mesh-tensorflow}). In both cases, model
parallelism is often used in conjunction with data parallelism~\cite{gpipe,
flexflow, mesh-tensorflow}, where each partition of the model is additionally
replicated across multiple accelerators, and the input batch is divided
evenly among these accelerators (Figure~\ref{fig:model-parallelism}, top).

The techniques presented in this paper can still be applied to this setting
to reduce resource requirements along the \textit{batch} dimension. More
specifically, within each model partition, the input batch can be divided
among virtual nodes rather than physical accelerators, and multiple virtual
nodes can be mapped to each accelerator as before. The system would effectively
unroll the data parallel pipelines into sequential forward and backward passes
(Figure~\ref{fig:model-parallelism}, bottom), thus trading off compute time for
lower resource requirement as in \name. This would bring the benefits of
reproducibility, elasticity and heterogeneous training to the model parallelism
setting as well. Exploring how to pipeline these virtual nodes for higher
efficiency (as in GPipe~\cite{gpipe} and PipeDream~\cite{pipedream}) would be
an interesting future direction.

\section{Related Work}
\label{sec:related}

\textbf{Gradient accumulation.} The execution of virtual nodes is
similar to gradient accumulation in PyTorch~\cite{pytorch-grad-accum}
and a variant of asynchronous training that synchronizes gradients
every $k$ steps~\cite{ksync, adacomm, elastic-bsp}. \name is a
generalization of these approaches: virtual nodes not only allow
users to simulate larger batch sizes, but also provide a general
abstraction that enable elasticity and heterogeneous training.

\textbf{Virtual nodes.} The use of virtual nodes to decouple from
hardware is not new. Chord~\cite{chord} uses virtual nodes to map
multiple ring segments to the same server, and Dynamo~\cite{dynamo}
uses virtual nodes to dynamically balance load across servers in the
system. \name borrows from these ideas.

\textbf{Elasticity mechanism.} Resource elasticity for deep learning
has been explored in~\cite{andrew-resource-elasticity}, and our implementation
builds on top of the resizing mechanisms they introduced. Elastic
Horovod~\cite{elastic-horovod} and TorchElastic~\cite{torchelastic}
also provide elasticity mechanisms for deep learning. Unlike \name,
however, these approaches do not provide model convergence guarantees
as they allow the global batch size to change throughout training.

\textbf{Cluster scheduling.} Multi-tenant GPU cluster schedulers such
as Optimus~\cite{optimus}, Tiresias~\cite{tiresias}, Gandiva~\cite{gandiva},
Themis~\cite{themis}, and Gavel~\cite{gavel} dynamically migrate jobs
across GPUs to achieve various cluster-level objectives. However, unlike
\name, these approaches constantly interrupt and restart jobs when adjusting
their resource allocations (\eg every 6 minutes~\cite{gavel}), and assume
fixed resource requirements for each job, thus limiting the scheduling
options available. Antman~\cite{antman} can adjust resource allocations
seamlessly by swapping to and from host memory, but only for co-located
jobs that share the same accelerators.

\textbf{Heterogeneous training.} Gavel~\cite{gavel} introduces policies
for heterogeneous clusters, but only makes homogeneous allocations.
Extending their scheduler to additionally consider \name's heterogeneous
training opens up new scheduling options, which can improve cluster
utilization (\S\ref{subsubsec:het-scheduler}).

\section{Conclusion}
\label{sec:conclusion}

\name is an important step towards decoupling deep learning
models from the underlying hardware. Leveraging the idea of
virtual node processing, \name allows users to reproduce
training results consistently across different clusters,
reap the benefits of resource elasticity without worrying
about model convergence, and combine multiple accelerator
types to improve training throughput, all without a single
change to the model specification or the hyperparameters.

The benefits of virtual nodes are not limited to the use cases
explored in this paper. In the future, we expect to see more
complexities associated with resource management pushed into
the deep learning frameworks themselves, enabling the user to
focus on application-level objectives instead.

\bibliographystyle{plain}
\bibliography{\jobname}

\end{document}